\documentclass[draft]{agujournal2019}
\usepackage{url} %this package should fix any errors with URLs in refs.
\usepackage{lineno}

\usepackage{booktabs}
\usepackage{colortbl}
\usepackage{amsmath}
\usepackage{MnSymbol}

\definecolor{color0}{rgb}{0.12, 0.47, 0.71}
\definecolor{color1}{rgb}{1.0, 0.5, 0.05}
\definecolor{color2}{rgb}{0.84, 0.15, 0.16}
\definecolor{color3}{rgb}{0.55, 0.34, 0.29}
\definecolor{color4}{rgb}{0.89, 0.47, 0.76}
\definecolor{color5}{rgb}{0.74, 0.74, 0.13}
\definecolor{color6}{rgb}{0.09, 0.75, 0.81}

\drafttrue

\journalname{JGR: Oceans}

\begin{document}

%% ------------------------------------------------------------------------ %%
%  Title
%
% (A title should be specific, informative, and brief. Use
% abbreviations only if they are defined in the abstract. Titles that
% start with general keywords then specific terms are optimised in
% searches)
%
%% ------------------------------------------------------------------------ %%

% Example: \title{This is a test title}

\title{The Fraction of Broken Waves in Natural Surf Zones}

%% ------------------------------------------------------------------------ %%
%
%  AUTHORS AND AFFILIATIONS
%
%% ------------------------------------------------------------------------ %%

% Authors are individuals who have significantly contributed to the
% research and preparation of the article. Group authors are allowed, if
% each author in the group is separately identified in an appendix.)

% List authors by first name or initial followed by last name and
% separated by commas. Use \affil{} to number affiliations, and
% \thanks{} for author notes.
% Additional author notes should be indicated with \thanks{} (for
% example, for current addresses).

% Example: \authors{A. B. Author\affil{1}\thanks{Current address, Antartica}, B. C. Author\affil{2,3}, and D. E.
% Author\affil{3,4}\thanks{Also funded by Monsanto.}}

\authors{C. E. Stringari\affil{1}, H. E. Power\affil{1}}

\affiliation{1}{University of Newcastle, School of Environmental and Life Sciences, Newcastle, Australia}
% \affiliation{2}{University of Queensland, School of Earth and Environmental Sciences, St Lucia, Australia}
% \affiliation{2}{Second Affiliation}
% \affiliation{3}{Third Affiliation}
% \affiliation{4}{Fourth Affiliation}

%(repeat as many times as is necessary)

%% Corresponding Author:
% Corresponding author mailing address and e-mail address:

% (include name and email addresses of the corresponding author.  More
% than one corresponding author is allowed in this LaTeX file and for
% publication; but only one corresponding author is allowed in our
% editorial system.)

% Example: \correspondingauthor{First and Last Name}{email@address.edu}

\correspondingauthor{C. E. Stringari}{Caio.EadiStringa@uon.edu.au}

%% Keypoints, final entry on title page.

%  List up to three key points (at least one is required)
%  Key Points summarize the main points and conclusions of the article
%  Each must be 100 characters or less with no special characters or punctuation

% Example:
% \begin{keypoints}
% \item	List up to three key points (at least one is required)
% \item	Key Points summarize the main points and conclusions of the article
% \item	Each must be 100 characters or less with no special characters or punctuation
% \end{keypoints}

\begin{keypoints}

\item The fraction of broken waves is a highly variable parameter in natural surf zones.

\item  There are links between the fraction of broken waves and surf zone processes such as infragravity waves and tides.

\item The probability distribution of broken wave heights is well described by a Weibull distribution.

\end{keypoints}

%% ------------------------------------------------------------------------ %%
%
%  ABSTRACT

%% ------------------------------------------------------------------------ %%

%% \begin{abstract} starts the second page

\begin{abstract}

This paper presents a novel quantification of the fraction of broken waves ($Q_b$) in natural surf zones using data from seven wave-dominated Australian beaches. $Q_b$ is a critical, but rarely quantified, parameter for parametric surf zone energy dissipation models which are commonly used as coastal management tools. Here, $Q_b$ is quantified using a combination of remote sensing and in-situ data. These data and machine learning techniques enable quantification of Qb for a substantial dataset ($>350,000$ waves). The results show that $Q_b$ is a highly variable parameter with a high degree of inter- and intra-beach variability. Such variance could be explained (at least partially) by correlations between $Q_b$ and environmental parameters. Tidal variations drive changes in $Q_b$ of up to 70\% for a given local water depth ($h$) on steep beaches, and increased infragravity energy levels decreased terminal values of $Q_b$ by about 20\%. The links between $Q_b$ and environmental forcing lead to the development of a correspondence between $Q_b$ and the Australian beach morphodynamic model. $Q_b$ is larger for a given normalized depth ($h/H_{m0_{\infty}}$, where $H_{m0_{\infty}}$ is offshore wave height) for dissipative beaches than for intermediate beaches. Finally, when comparing data to existing models, three commonly used theoretical formulations for $Q_b$ are observed to be poor predictors with errors of the order of 40\%. Existing theoretical Qb models are shown to improve (revised errors of the order of 10\%) if the Rayleigh probability distribution that describes the wave height is in these models is replaced by the Weibull distribution. 

\end{abstract}

\section{Introduction}\label{intro}

As waves approach shallow water, they undergo transformations where energy is dissipated via a plethora  of phenomena, with wave breaking being the most significant. On gently sloping beaches, the surf zone is wide enough for most of the wave energy to be dissipated. In contrast, on steep beaches, the surf zone is narrow and the incoming energy is reflected with little or no breaking, or drives swash processes \cite{Wright1984, Wright1982}. The most common way of tracking the energy dissipation is through the cross-shore variation in  wave height decay. Historically, wave heights in the surf zone have been modelled using probabilistic or parametric approaches \cite{Baldock1998}. Probabilistic models track wave height decay by propagating each component of a joint probability distribution of wave heights ($H$) and periods ($T$) shoreward. This approach frequently implies saturation, i.e., the wave height in shallow water is only a function of the local water depth ($h$) and of the wave height to water depth ratio, or breaker index, $\gamma$, such that $H=h*\gamma$ \cite{Thornton1982, HajimeMaseandYuichiIwagaki, Dally1986, Horikawa1966}. Parametric models, on the other hand, use an energy balance between the incoming energy flux and the local energy dissipation to track wave height decay  \cite{LeMehaute1962}:

\begin{equation}
    \frac{\partial E C_{g}}{\partial{x}} = - \langle \epsilon \rangle \label{E}
\end{equation}

in which $E$ is the incoming energy in the cross-shore direction ($x$, positive onshore), $C_g$ is the wave group speed, and $ \langle \epsilon \rangle$ is the combined time-averaged energy dissipation due to wave breaking and bottom friction. \citeA{Thornton1983} showed that dissipation via bottom friction on sandy beaches is small and, therefore, can be neglected in the dissipation term. Thus,  leaving $\langle \epsilon \rangle$ to represent dissipation due to bottom induced wave breaking. The energy dissipation is then derived from a bore model \cite{LeMehaute1962, Battjes1978}:

\begin{equation}
    \epsilon = \frac{1}{4} \rho g f_{p} B_f \frac{H^3}{h} Q_b \label{E2}
\end{equation}

where $\rho$ is the water density, $f_p$ is the wave peak frequency, $g$ is the gravitational acceleration, $B_f$ is a free parameter of order 1 (O(1)), and $Q_b$ is the fraction of broken waves at any given point in the surf zone.

% - determining the energy dissipation in the surf zone and, therefore, directly controlling how the model adapts to changes in water depth \cite{Battjes1978} -

Several parametric models based on Equations \ref{E} and \ref{E2} have been developed to describe wave energy dissipation and to predict wave heights in the surf zone. These models are computationally efficient, relatively simple to use, and are often used alongside other applications as coastal management tools \cite{Baldock1998, Ruessink2003, Apotsos2008, Alsina2007, Janssen2007}. The main difference between parametric models is how the dissipation term ($\epsilon$) is evaluated. From Equation \ref{E2}, it follows that $\epsilon$ is a direct function of the fraction of broken waves ($Q_b$) \cite{LeMehaute1962, Battjes1978}. Although $Q_b$ is a critical parameter for the model, very few authors have directly quantified $Q_b$, or compared it to models predictions using data from natural beaches. To the author's knowledge, only \citeA{Thornton1983} and, more recently, \citeA{Carini2015} have made attempts to do this.

Despite the vast literature on parametric wave modelling, there remain several unanswered questions. Firstly, no probability density function (PDF or $p(\underbar{x})$, where $\underbar{x}$ is a random variable) to describe broken waves is known besides the approximations of \citeA{Thornton1983}. Such approximations are not mathematically true PDFs and, as observed by \citeA{Baldock1998}, may not generalise to all beach types and slopes. Secondly, the extent to which errors in $Q_b$ affect the full dissipation in the models after they are individually optimised for specific beaches are currently unknown \cite{Apotsos2008}. Lastly, no links between $Q_b$ and other surf zone processes have been clearly established. The primary objective of this paper is to address these three knowledge gaps.

In this paper, two novel techniques to obtain $Q_b$ are presented. Firstly, by directly obtaining it from collocated video imagery and pressure transducer (PT) records, and secondly, by using state-of-the-art machine learning and artificial intelligence methods. Comparison of field data to the available analytical models of $Q_b$ is conducted and novel links between $Q_b$ and infragravity waves, tides, and beach morphodynamics are presented. This paper is organised as follows. Section \ref{QB_Theory_01} reviews the mathematical formulations of $Q_b$. Section \ref{methods} describes the data collection, data pre-processing, and presents direct and indirect methods to quantify $Q_b$. Section \ref{results} presents the results of field observations, establishes links between $Q_b$ and surf zone parameters, and provides a novel description of the PDF of broken wave heights. Finally, the results are discussed in Section \ref{discussion}, and conclusions are provided in Section \ref{conclusion}.

\section{Review of Theoretical Formulations of $Q_b$}\label{QB_Theory_01}

\begin{figure}[htp]
  \centering
  \includegraphics[width=0.95\textwidth]{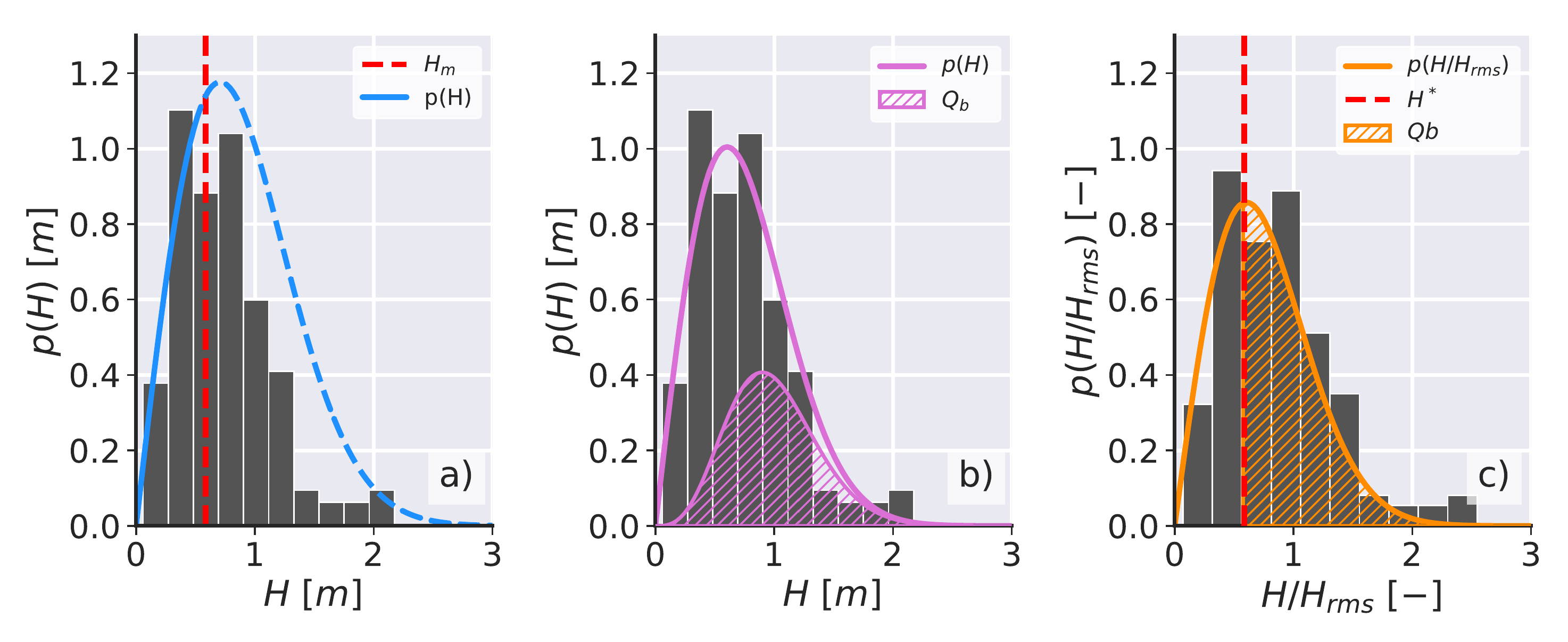}
  \caption{Illustration of the theoretical formulations for $p(H)$ and $Q_b$. a) BJ78. For this particular case, $\hat{H}$=$0.6$ and $H_{max}$=$0.75$. b) TG83 formulation for the same Rayleigh PDF as in a). The hatched area represents Equation \ref{QB_TG1983}. In this example, $\gamma$=$0.42$, $h$=$1.2$ and $n$=$2$. c) B98 formulation for the same Rayleigh PDF as in a). In this example, $H^*$=$0.6$, and the dashed area represents Equation \ref{QB_B98}. In all panels, the histograms were drawn from randomly generated wave heights obtained from the Rayleigh PDF shown in a). Note that $Q_b$ cannot be obtained geometrically for BJ78.}\label{qb_plot}
\end{figure}

There are three main formulations for estimating $Q_b$ analytically, all of which rely on obtaining an expression derived from PDFs of the wave height ($p(H)$). The first approach, described in \citeA{Battjes1978}, hereafter BJ78, uses a truncated Rayleigh PDF to approximate $Q_b$ (Figure \ref{qb_plot}-a). In this formulation, all waves higher than a threshold wave height ($H_{max}$) are considered to be breaking such that:

\begin{align}
 p(\underbar{H} \leq H) & = 1 - \exp\left( -\frac{1}{2}  \left( \frac{H}{\hat{H}} \right)^2 \right) & for \ 0 \leq H \leq H_{max}  \nonumber \\
      & = 1 & for \ H_{max} \leq H
\end{align}

in which $\hat{H}$ is the modal value that defines the Rayleigh $PDF$ for $H$, $\underbar{H}$ is any given random wave height, and $H_{max}$ is the maximum wave height possible for a given sea-state. From this expression, it is possible to obtain an explicit equation for $Q_b$ \cite{Battjes1978}:

\begin{equation}
    Q_b = p \left( \underbar{H} = H_{max} \right) = \exp{\left( -\frac{1}{2} \left( \frac{H_{max}}{\hat{H}} \right)^2 \right)}. \label{QB_B1978a}
\end{equation}

Equation \ref{QB_B1978a} can be re-written as an implicit relation  considering that $\hat{H}$ is equivalent to $H_{rms}$:

\begin{equation}
	\frac{1-Q_b}{\ln{Q_b}} = \frac{H_{rms}}{H_{max}} \label{QB_B1978}
\end{equation}

in which $H_{rms}$ is the root-mean-square wave height, and $H_{max}$ can be obtained from \citeA{Miche1934} formulation for the maximum wave height, or from any other compatible formulation. BJ78 observed that their model underestimated energy dissipation closer to the shore and adopted $H_{rms} = H_{max}$ if $H_{rms}>H_{max}$, i.e., a saturated surf zone. This assumption has been shown to not fully account for the energy dissipation, particularly for steeper beach profiles \cite{Baldock1998, Alsina2007, Janssen2007}. In addition, surf has been shown to be unsaturated on multiple beaches \cite{Power2010}. Nonetheless, Equation \ref{QB_B1978} is frequently used in its spectral form \cite{Eldeberky1996} in wave forecasting models such as SWAN and WW3 \cite{holthuijsen2010, WW32016}.

\citeA{Thornton1983}, hereafter TG83, used a similar approach to model $Q_b$ but instead of using a truncated Rayleigh PDF, they observed that a full Rayleigh PDF better described their field data:

\begin{equation}
    p(H) = \frac{2H}{{H_{rms}}^2} \exp{ \left[ - \left( \frac{H}{H_{rms}} \right)^2 \right] }. \label{pH_TG1983}
\end{equation}

$Q_b$ was then obtained by integrating empirical equations designed to describe the probability of broken waves ($p_b(H)$, dashed area in Figure \ref{qb_plot}-b)

\begin{equation}
	Q_b = \int^{\infty}_{0} p_b(H)dH \leq 1 \label{QB_TG1983}
\end{equation}

where

\begin{equation}
	p_b(H) = W(H)*p(H) \label{QB_TG1983a}
\end{equation}

and,

\begin{equation}
	W(H) = \left( \frac{H_{rms}}{\gamma h} \right)^n * \mathcal{F} \label{QB_TG1983b}
\end{equation}

in which $\gamma$ is the wave height to water depth ratio, $h$ is the averaged local water depth, $n$ is a free parameter, and $\mathcal{F}$ is a scaling factor that was used to give more weight to higher waves. Note that Equation \ref{QB_TG1983a} is not a true PDF because, if not bounded, it can result in $\displaystyle \int_{0}^{\infty} p_b(H)dH \neq 1$. In their study, TG83 used, $\gamma=0.42$, $n=\{2, 4\}$ and

\begin{equation}
	\mathcal{F} = 1 -\exp{ \left[ - \left( \frac{H}{\gamma h} \right)^2 \right] }.
\end{equation}

\citeA{Baldock1998}, hereafter B98, followed TG83 and also used a full Rayleigh PDF to describe the wave height distribution in the surf zone. In their case, $H$ was normalised by $H_{rms}$ such that:

\begin{equation}
    p\left( \frac{H}{H_{rms}} \right) = 2\frac{H}{H_{rms}} \exp{ \left[ - \left( \frac{H}{H_{rms}} \right)^2 \right] } \label{pH_B98}
\end{equation}

Given the observation in B98 that TG83's empirical formulations for $p_b(H)$ may not be universal for all surf zone conditions and beach slopes, B98 formulated $Q_b$ so that it is obtained by integrating Equation \ref{pH_B98} for all waves in which $H/H_{rms} \geq H_b/H_{rms}$ (hatched area in Figure \ref{qb_plot}-c):

\begin{equation}
    Q_b = \int^{\infty}_{H^{*}} p\left( \frac{H}{H_{rms}} \right) d \left( \frac{H}{H_{rms}} \right) \ = \ \exp{ \left[ - \left( \frac{H_b}{H_{rms}} \right)^2 \right] } \label{QB_B98}
\end{equation}

where ${H^{*}}$ is the limiting parameter $H_b/H_{rms}$, and $H_b$ can be obtained from any formulation for the maximum breaker height, for example as per \citeA{Battjes1985}:

\begin{equation}
    \frac{H_b}{h} = 0.39 + 0.56\tanh(33S_{\infty})
    \label{Hb}
\end{equation}

where $S_{\infty}$ is the offshore wave steepness. This definition of $H_b$ is used hereafter a maximum breaker height, or a wave height to water depth ratio ($H_b/h=\gamma$). Note, however, that such definitions of $\gamma$ and $H_b/h$ are not necessary equivalent \cite{Power2010, Raubenheimer1996}. This approach was only undertaken so that any errors this formation (Equation \ref{Hb}) may contain are shared between the models and do not influence in the comparisons between them.  There are, nonetheless, other formulations for $\gamma$ that can  produce significantly different results for the energy dissipation when adapted into the formulations for $Q_b$ (see Section \ref{model_disc}) \cite{Apotsos2008, Ruessink2003}; however, investigating these is beyond the scope of this study.

\subsection{$Q_b$ Based on a Weibull PDF}\label{QB_Theory_02}

A feature that all the formulations for $Q_b$ share is the Rayleigh PDF from which $Q_b$ is obtained. Several authors have proposed that either a full or a modified  Weibull PDF is a better descriptor of $p(H)$, especially in shallower water depths \cite{Mase1989,Hameed1985,Battjes2000,Mendez2004,Power2016}. For non-negative values of $H$, $p(H)$ is described by a Weibull PDF as:

\begin{equation}
    p(H) = \frac{\kappa}{H_{rms}} \left( \frac{H}{H_{rms}} \right)^{\kappa-1} \exp{\left[- \left( \frac{H}{H_{rms}} \right)^{\kappa} \right] } \label{pH_Weibull}
\end{equation}

in which $\kappa$ defines the shape of the distribution and can be any positive number, and $H_{rms}$ is the scale parameter of the distribution.  For shape parameter $\kappa=2$, the Weibull PDF reduces to the Rayleigh PDF. Following TG83, Equation \ref{pH_Weibull} can be used to obtain $Q_b$ in the same manner as in Equation \ref{QB_TG1983a}:

\begin{equation}
    Q_b = \int^{\infty}_{0} p_{b}(H) dH \leq 1 \label{QB_TG1983_w}
\end{equation}

with the probability of broken waves being

\begin{equation}
     p_{b}(H) = W(H)*p(H) \label{QB_TG1983_w_a}
\end{equation}

where

\begin{equation}
    W(H) = \left( \frac{H_{rms}}{\gamma h} \right)^{n} * \left( 1 -\exp{ \left[ - \left( \frac{H}{\gamma h} \right)^{\kappa} \right] } \right). \label{QB_TG1983_w_b}
\end{equation}

Similarly, it is possible to modify B98 to use the Weibull PDF:

\begin{equation}
    p \left( \frac{H}{H_{rms}}  \right) = \kappa \left(\frac{H}{H_{rms}} \right)^{\kappa-1} \exp{\left[- \left( \frac{H}{H_{rms}} \right)^{\kappa} \right] }. \label{pH_B98_W}
\end{equation}

Such that $Qb$ can be obtained by integrating Equation \ref{pH_B98_W} for all waves greater than $H^*$:

\begin{equation}
    Q_b = \int^{\infty}_{H^{*}} p\left( \frac{H}{H_{rms}} \right) d \left( \frac{H}{H_{rms}} \right) \ = \ \exp{\left[- \left( \frac{H_b}{H_{rms}} \right)^{\kappa} \right] }. \label{QB_B98_w}
\end{equation}

\begin{figure}[htp]
  \centering
  \includegraphics[width=0.95\textwidth]{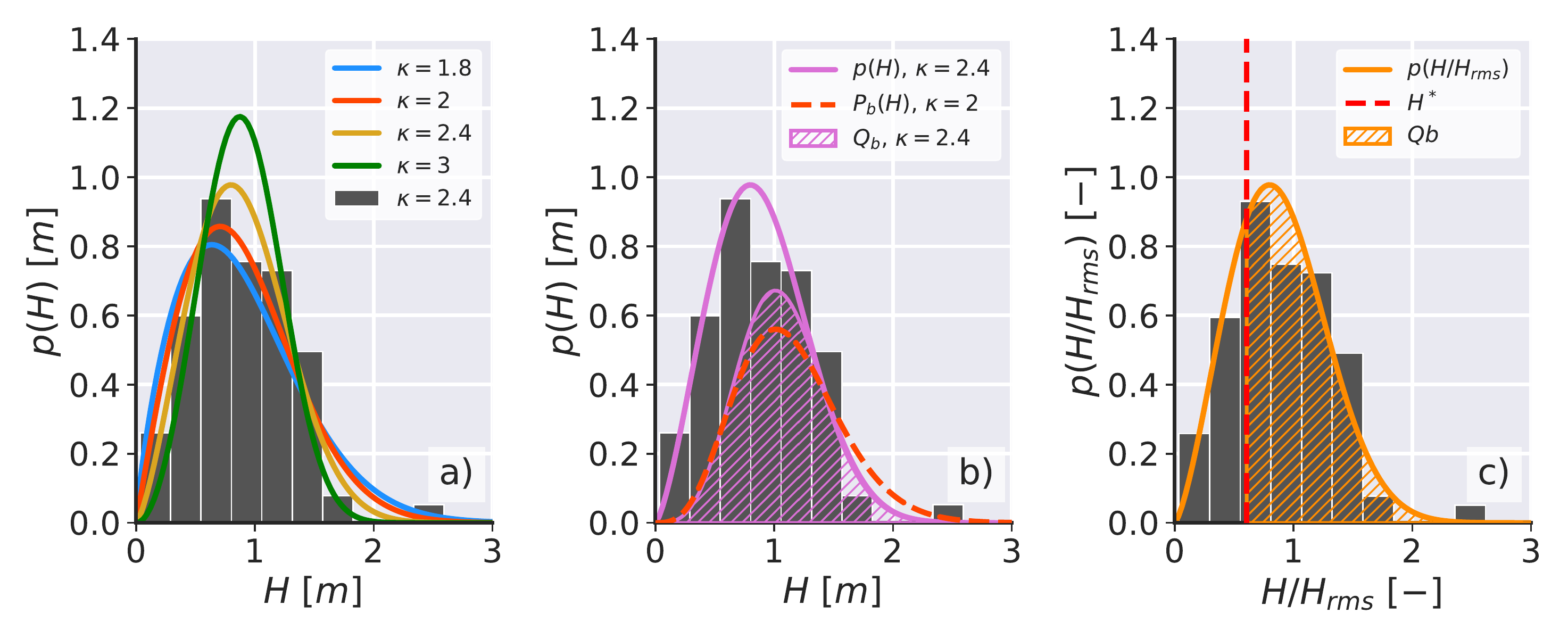}
  \caption{Illustration of the adapted theoretical formulations for $p(H)$ and $Q_b$. a) The Weibull PDF for different values of $\kappa$. In this example, $H/H_{rms}$=$1$. b) Adapted TG83 formulation for the Weibull PDF. The hatched area represents Equation \ref{QB_TG1983_w_a}. In this example, $\gamma$=$0.42$, $h$=$1.23$ and $n$=$2$. c) Adapted B98 formulation for the Weibull PDF. In this example, $H^*$=$0.6$. The dashed area represents Equation \ref{QB_B98_w}. In all plots, the histograms were drawn from random wave heights obtained from the Weibull PDF with $\kappa$=$2.4$.}\label{qb_plot_new}
\end{figure}

Figure \ref{qb_plot_new}-a shows a graphical representation of the Weibull PDF for various values of $\kappa$, and how $Q_b$ can be obtained from Equation \ref{QB_TG1983_w} (Figure \ref{qb_plot_new}-b) and Equation \ref{QB_B98_w} (Figure \ref{qb_plot_new}-c). The Weibull PDF with $\kappa$ $>$ $2$ should automatically increase $Q_b$ in TG83's model if the exponent $n$ in the scaling factor $\mathcal{F}$ is kept unchanged from the original values $\{2, 4\}$ (note the orange dashed line in Figure \ref{qb_plot_new}-b). The major issue with the use of this alternative PDF is the inclusion of a new free parameter ($\kappa$) that needs to be obtained from the available data. In this paper, the optimal value of $\kappa=2.4$ found by \citeA{Power2016} using an extensive natural surf zone dataset is used, and $n$ is set to 1 when TG83's model is adapted to use the Weibull PDF.

\section{Methods}\label{methods}

\subsection{Field Data Collection}\label{data_collection}

\begin{figure}[htp]
  \centering
  \includegraphics[width=0.95\textwidth]{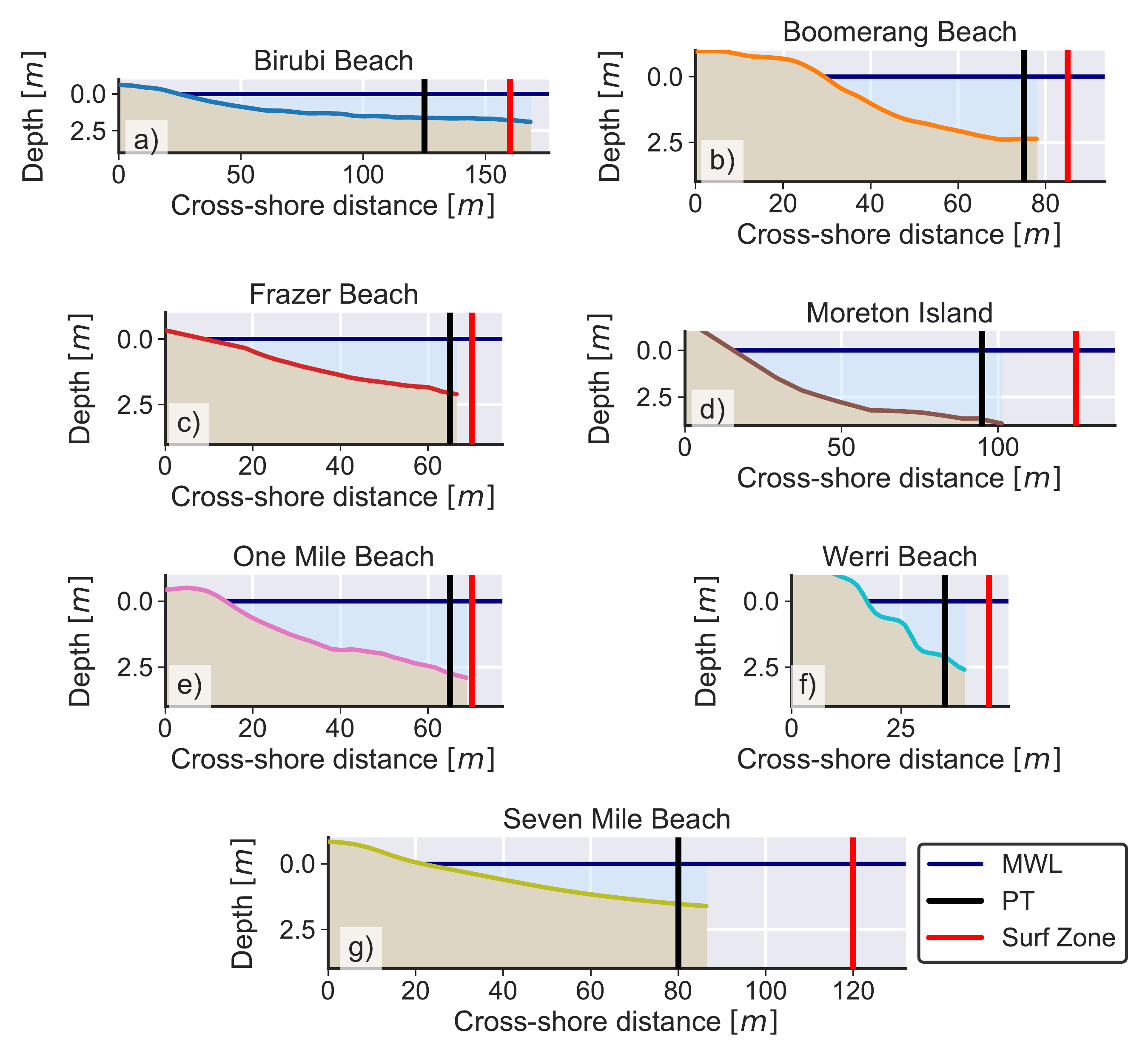}
  \caption{Representative beach profiles for each location. The thick black line shows the location of the offshore most PT, the thick red line shows the averaged location of the surf zone limit observed from the video data, and the dark blue line indicates the mean water level (MWL) calculated based on the offshore-most PT over the duration of the experiment. a) Birubi beach, New South Wales (NSW) (05/07/2017), b) Boomerang Beach, NSW (20/09/2019), c) Frazer Beach, NSW (24/04/2018), d) Moreton Island, Queensland (20/12/2016), e) One Mile Beach (Foster), NSW (07/08/2014), f) Seven Mile Beach (Kiama), NSW (13/08/2014), and g) Werri Beach (16/08/2014). All the panels have the same axes aspect ratio (6).}\label{profiles}
\end{figure}

PT data and video imagery of the nearshore were collected at seven different sandy micro-tidal, wave-dominated Australian beaches during 19 individual deployments (Table \ref{expt_table} and Figure \ref{profiles}). Each experiment consisted of deploying PTs attached to a chain that sat on the seabed and extended in a cross-shore orientation. This was combined with collecting video imagery of the surf zone from an elevated location (headland or house balcony) over an individual tidal cycle. The PTs were always deployed at the seabed level and covered cross-shore extents ranging from 30 to 120m (from the beach face) on surf zones of about the same width. Therefore, the vast majority of the collected data was from the outer and inner surf zones. The PTs used in this study were either INW PT2X (8$Hz$ or 10$Hz$) or RBR Solo (16$Hz$) programmed to record at a minimum sampling rate of 8$Hz$. The video cameras were consumer-grade Sony cameras attached to a surveying tripod (Sony HDR-XR200 for the 2014 experiments and Sony-HDR-CX240 for the remainder of the experiments). Video cameras were calibrated to compensate for lens distortions as per \citeA{Holland1997}.

In addition to the PT and video data collection, all the beaches were surveyed at low tide, and a minimum number of four ground control points (GCPs) and one beach profile were acquired. Beach profiles covered the foredune to the maximum depth allowed by environmental conditions. The beach slopes ($\beta$) were calculated from the surf-swash boundary to the seaward-most surveyed PT and should, therefore, be representative for the surf zone (see Figure \ref{profiles}). The spectral surf zone parameters in Table \ref{expt_table}  were calculated following \citeA{holthuijsen2010} and the offshore data were obtained from the New South Wales wave transformation toolbox \cite{NSWOfficeofEnvironmentandHeritage} except for Moreton Island for which offshore data were obtained from the Brisbane wave-rider buoy \cite{QueenslandGovernment-ScienceEnvironment2018}.

The beaches in this study covered the full range of morphodynamic states \cite{Wright1984} except for reflective beaches due to the absence of a surf zone in this beach type, which precludes obtaining meaningful $Q_b$ data. The northern end of Seven Mile Beach (Gerroa, New South Wales) (Figure \ref{profiles}-f) represented the dissipative state, Birubi Beach (Figure \ref{profiles}-a) represented the alongshore bar and trough state(LBT), Boomerang Beach (Figure \ref{profiles}-b) and Moreton Island  (Figure \ref{profiles}-d) represented the rhythmic bar and beach state (RBB),  One Mile Beach (Figure \ref{profiles}-e), Frazer Beach (Figure \ref{profiles}-c) and Boomerang Beach represented the transverse bar and rip state (TBR) and, Werri Beach (Figure \ref{profiles}-g) and One Mile Beach (Figure \ref{profiles}-e) represented the low tide terrace state (LTT). Some beaches presented different morphodynamic states at the same time, e.g., sections of Boomerang Beach were characterised as RBB (19/09/2016) whereas others presented TBR (20/09/2016 onward) morphology. However, note that the profiles shown in Figure \ref{profiles} may not fully illustrate the described classification because they are a two-dimensional representation of a three-dimensional system. The presented classification was based on Timex images \cite{Holman2007} of the beaches and  additional information from \citeA{short1999beaches, short2007beaches} (see Appendix 1).

\begin{table}
\caption{Data record for each experiment. $nPTs$ is the number of PTs in each deployment. $H_{m_0}$ and $T_{m_0}$ are the averaged surf zone significant wave height and significant wave period. $H_{{m_0}_{\infty}}$, $T_{{m_0}_{\infty}}$, and $Dp_{\infty}$ are the offshore significant wave height, significant wave period, and peak wave direction. $\beta$ is the beach gradient. $M. \ state$ is the observed morphodynamic state \cite{Wright1984}. Marker indicates the representation for each deployment used hereafter. The asterisk indicates collocated PT and video data used in the training step of the machine learning models.}\label{expt_table}
\resizebox{\textwidth}{!}{%
\begin{tabular}{lllrrrrrlrlc}
\toprule
{} &          Location &              Date &  $nPTs$ &  $H_{m_0}$ &  $T_{m_0}$ &  $H_{{m_0}_{\infty}}$ &  $T_{{m_0}_{\infty}}$ & $Dp_{\infty}$ &  $\beta$ & $M. \ state$ &                                       Marker \\
\midrule
1  &      Birubi Beach &        2017-07-05 &       9 &      0.330 &      7.843 &                 0.589 &                15.719 &             S &    0.011 &          LBT &         \LARGE \textcolor{color0}{$\bullet$} \\
2  &      Birubi Beach &  2017-07-06$^{*}$ &       9 &      0.378 &     12.310 &                 0.641 &                12.188 &             S &    0.007 &          LBT &    \LARGE \textcolor{color0}{$\blacksquare$} \\
3  &   Boomerang Beach &  2016-09-19$^{*}$ &       8 &      0.778 &      9.887 &                 2.027 &                 7.182 &             W &    0.066 &      RBB/TBR &         \LARGE \textcolor{color1}{$\bullet$} \\
4  &   Boomerang Beach &        2016-09-20 &       5 &      0.685 &      8.104 &                 1.275 &                 7.889 &             S &    0.085 &      RBB/TBR &    \LARGE \textcolor{color1}{$\blacksquare$} \\
5  &   Boomerang Beach &        2016-09-21 &       9 &      0.721 &      7.844 &                 1.213 &                 7.632 &            SE &    0.048 &      RBB/TBR &   \LARGE \textcolor{color1}{$\blacklozenge$} \\
6  &   Boomerang Beach &        2016-09-22 &      10 &      0.594 &      7.894 &                 1.685 &                 6.061 &           WSW &    0.049 &      RBB/TBR &  \LARGE \textcolor{color1}{$\blacktriangle$} \\
7  &      Frazer Beach &  2018-04-24$^{*}$ &      11 &      0.425 &      6.586 &                 1.750 &                 7.623 &            SE &    0.035 &          TBR &         \LARGE \textcolor{color2}{$\bullet$} \\
8  &    Moreton Island &        2016-12-19 &       7 &      0.665 &      9.175 &                 2.131 &                 9.979 &            SE &    0.069 &          RTB &         \LARGE \textcolor{color3}{$\bullet$} \\
9  &    Moreton Island &  2016-12-20$^{*}$ &      14 &      0.822 &      7.923 &                 1.462 &                10.302 &            SE &    0.069 &          RTB &    \LARGE \textcolor{color3}{$\blacksquare$} \\
10 &    One Mile Beach &        2014-08-04 &      13 &      0.669 &      8.868 &                 1.457 &                 8.825 &             S &    0.045 &      TBR/LTT &         \LARGE \textcolor{color4}{$\bullet$} \\
11 &    One Mile Beach &        2014-08-05 &      12 &      0.645 &      8.311 &                 1.188 &                 9.461 &            SE &    0.076 &      TBR/LTT &    \LARGE \textcolor{color4}{$\blacksquare$} \\
12 &    One Mile Beach &        2014-08-06 &      11 &      0.767 &     11.055 &                 1.464 &                11.491 &             S &    0.050 &      TBR/LTT &   \LARGE \textcolor{color4}{$\blacklozenge$} \\
13 &    One Mile Beach &  2014-08-07$^{*}$ &       9 &      0.734 &     10.673 &                 1.638 &                 8.892 &             S &    0.048 &      TBR/LTT &  \LARGE \textcolor{color4}{$\blacktriangle$} \\
14 &  Seven Mile Beach &  2014-08-13$^{*}$ &       7 &      0.585 &     10.066 &                 1.565 &                12.702 &           SSE &    0.037 &            D &         \LARGE \textcolor{color5}{$\bullet$} \\
15 &  Seven Mile Beach &        2014-08-14 &       9 &      0.556 &      9.527 &                 0.964 &                11.495 &           SSE &    0.035 &            D &    \LARGE \textcolor{color5}{$\blacksquare$} \\
16 &  Seven Mile Beach &        2014-08-19 &       8 &      0.765 &     12.937 &                 1.140 &                12.000 &           SSE &    0.028 &            D &   \LARGE \textcolor{color5}{$\blacklozenge$} \\
17 &  Seven Mile Beach &        2014-08-20 &       8 &      0.630 &     11.648 &                 1.043 &                10.133 &           SSE &    0.029 &            D &  \LARGE \textcolor{color5}{$\blacktriangle$} \\
18 &       Werri Beach &        2014-08-15 &       9 &      0.959 &     10.089 &                 1.519 &                11.652 &            SE &    0.220 &          LTT &         \LARGE \textcolor{color6}{$\bullet$} \\
19 &       Werri Beach &  2014-08-16$^{*}$ &       8 &      0.956 &      9.278 &                 1.557 &                12.610 &           ESE &    0.114 &          LTT &    \LARGE \textcolor{color6}{$\blacksquare$} \\
\bottomrule

\end{tabular}
}
\end{table}

\subsection{Data Pre-processing}

\subsubsection{Wave-by-wave Analysis}\label{wave-by-wave}

\begin{figure}[htp]
 \centering
 \includegraphics[width=0.95\textwidth]{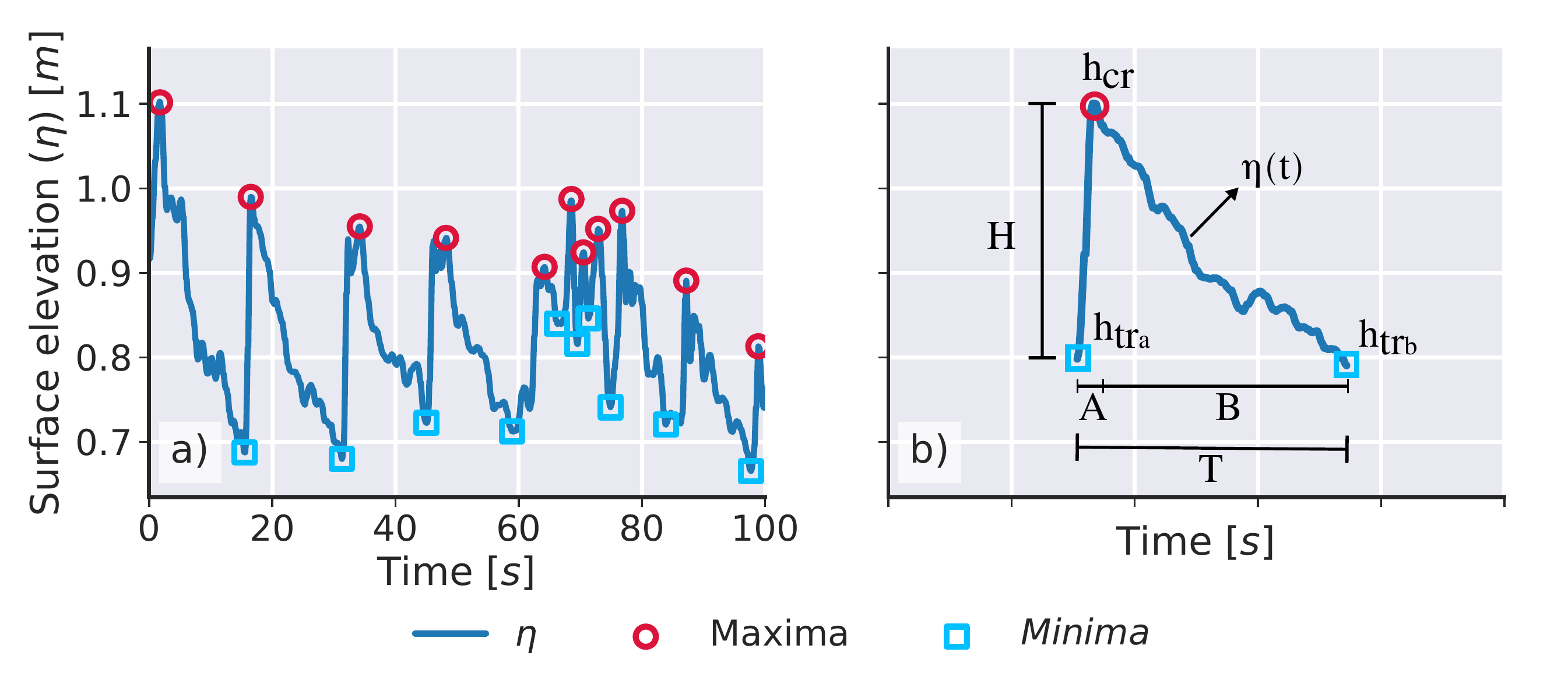}
 \caption{a) Example of the wave-by-wave analysis for Birubi Beach. The local maxima and minima are represented by the red and blue square markers, respectively. b) Schematic representation of the wave parameters in Table \ref{Xvector} (not to scale).}\label{wbw_plot}
\end{figure}

PT data were processed following \citeA{Power2010}. For each PT, the data record was divided into 15 minutes intervals to ensure stationarity with respect to the tide, and checked visually to ensure that there were no dry periods, i.e., to ensure that the  PT was not in the swash zone. Each of these time-series was re-sampled to 8$Hz$ (if needed) and individual waves were extracted using a wave-by-wave algorithm that searches for local minima and maxima in the series. Although the method is similar in concept to \citeA{Power2010}, the implementations did not share code. The present algorithm has two free parameters: a wave height threshold and a searching window. For this study, the searching window was set to two times the sampling frequency ($f$) and the wave height threshold was 15\% of the spectral significant wave height $H_{{m_0}}$ \cite{holthuijsen2010}. Obtaining the wave height threshold was somewhat subjective and the present threshold was chosen after visual inspection of several 15-minute time-series. No frequency filters were applied to the data to preserve the infragravity wave signal that is usually lost when using the more common combination of filtering and the zero-crossings method (e.g., as done in \citeA{Haller2009, Postacchini2014}). Figure \ref{wbw_plot} shows an example of the results of the wave-by-wave analysis applied to Birubi Beach. Even under the noisy conditions of this deployment, the algorithm performed well (Figure \ref{wbw_plot}-a).

\subsubsection{Video Imagery processing}

\begin{figure}[htp]
 \centering
 \includegraphics[width=0.95\textwidth]{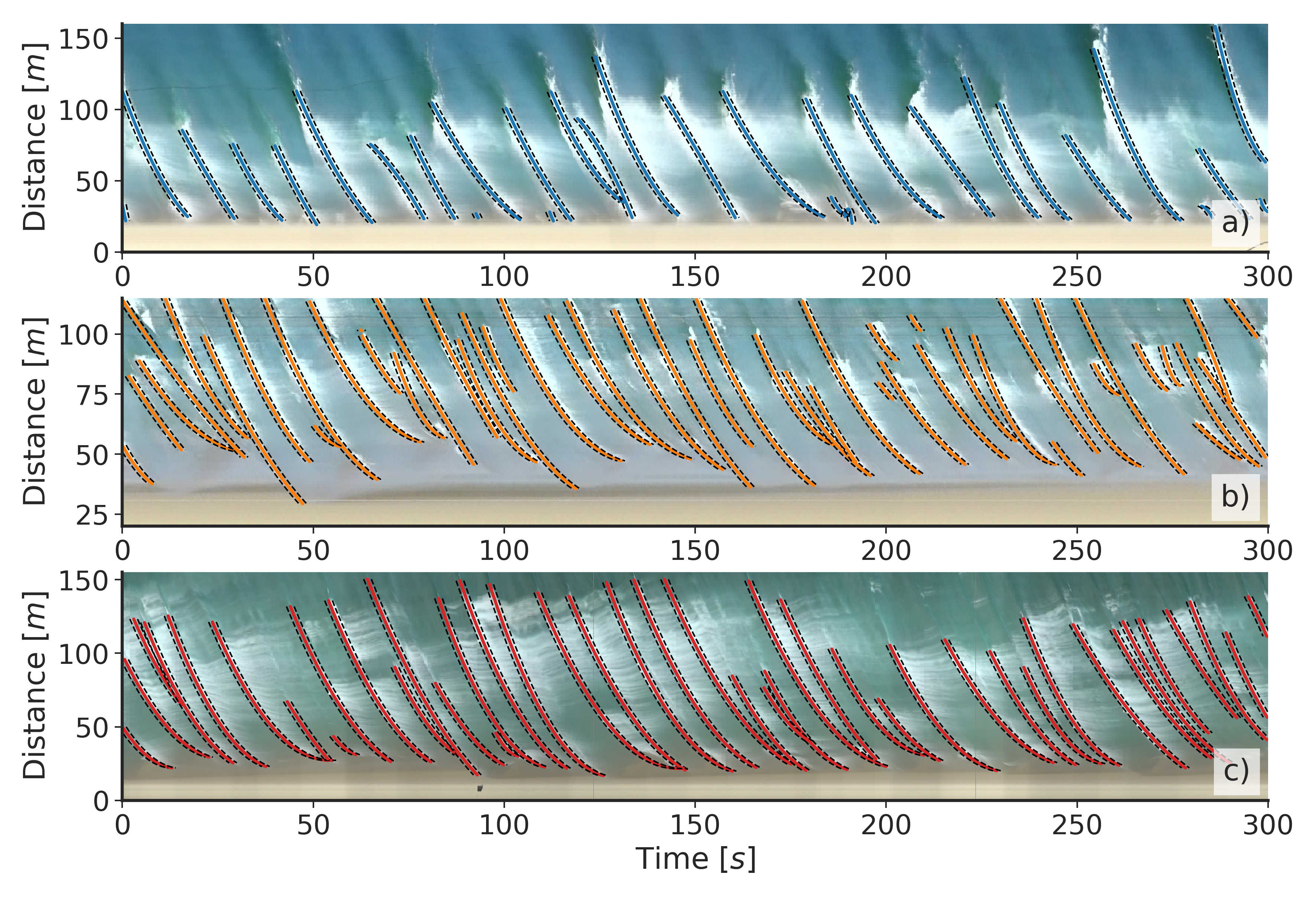}
 \caption{Example of timestacks and tracked wave paths for a) Boomerang Beach, b) Frazer Beach, and c) Moreton Island. The thick lines indicate the optimal wave path and the dashed black lines the confidence interval for each wave.}\label{timestacks}
\end{figure}

Video imagery data were processed following \citeA{Holland1997}. For each deployment, video data were down-sampled from $25Hz$ to $10Hz$, and individual frames were extracted. These frames were then rectified using the algorithm provided by \citeA{Hoonhout2015}. From each frame, a cross-shore array of pixels were extracted at the same location as the PT array and stacked in time, resulting in a timestack image \cite{Aagaard2016}.  The image coordinate system was translated and rotated so that it aligned with the beach profile and PT array (with the cross-shore coordinate oriented offshore). All timestacks were linearly interpolated with sub-pixel accuracy to a cross-shore resolution of $0.1m$. For each timestack, visible waves were tracked following \citeA{Stringari2019}, resulting in a collection of tracked wave paths for each location. Figure \ref{timestacks} shows representative timestacks and examples of tracked waves for Boomerang Beach, Frazer Beach, and Moreton Island. Examples for the remaining locations can be found in \citeA{Stringari2019}.

\subsection{Quantifying the Fraction of Broken Waves}\label{qb_quant}

\begin{figure}[htp]
 \centering
 \includegraphics[width=0.95\textwidth]{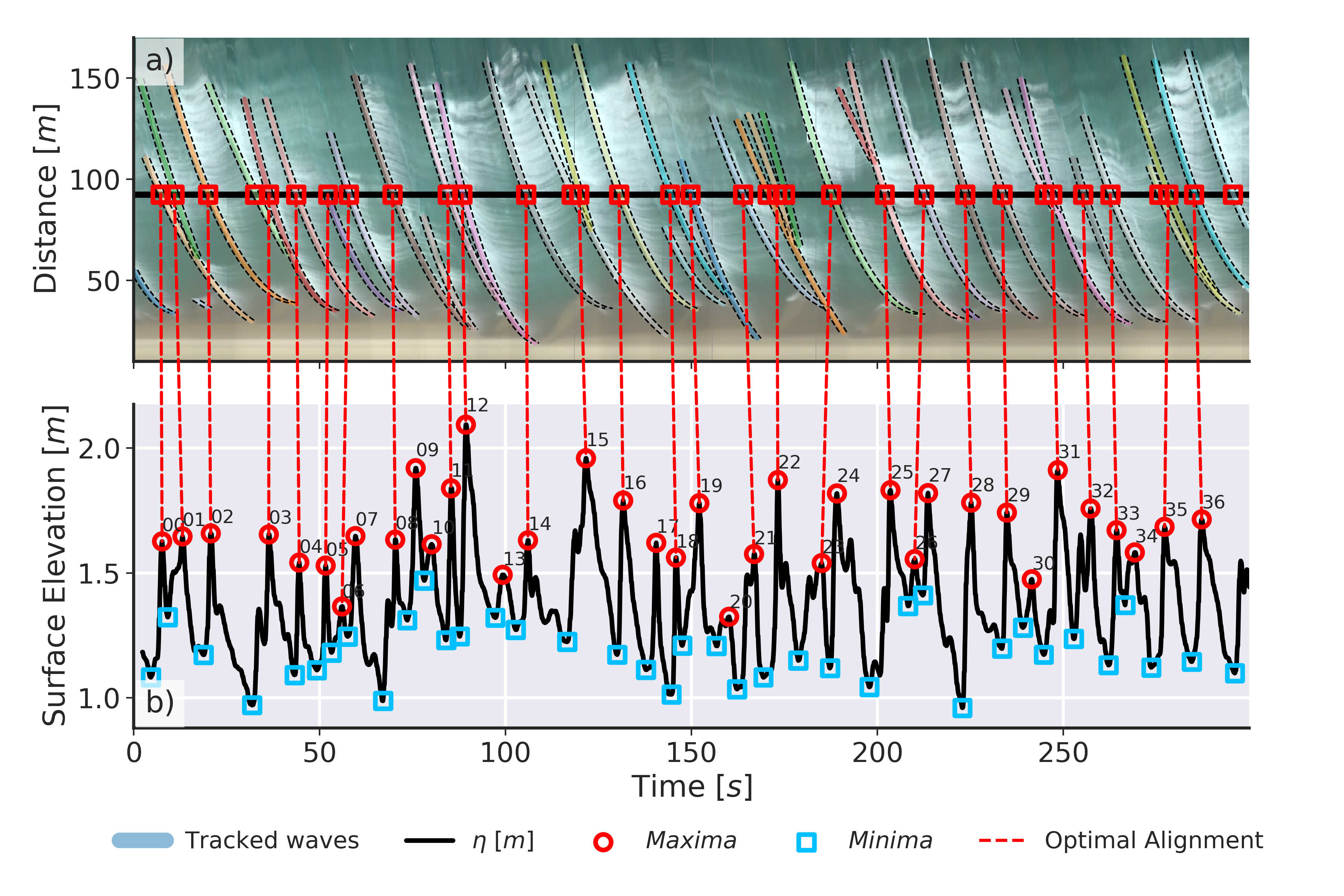}
 \caption{Example of 5 minutes of data alignment. a) timestack and tracked wave paths (coloured lines) and confidence intervals (dashed lines) for Moreton Island. The thick black line indicates the cross-shore location of one PT in the surf zone. b) Water surface elevation record at the cross-shore location shown in a). The individual waves were obtained via the wave-by-wave analysis outlined in Section \ref{wave-by-wave}. The vertical connection lines indicate the optimal alignment between the two data. In this particular example $Q_b$=$0.73$.  }\label{qb_collocated}
\end{figure}

The collocated PT and video data allowed for the direct quantification of $Q_b$. This was done by aligning the tracked waves with the corresponding PT data in both space and time (Figure \ref{qb_collocated}). The cross-shore spatial alignment was done directly from surveyed data, and the alignment in time was done by shifting the PT time-series by an optimal time delay. This delay was obtained via cross-spectral correlation between the pixel intensity time-series and the water surface elevation ($\eta$) at a given location (see \citeA{Stringari2019} for more details). Following the time-series alignment, each wave crest in the PT record  was attempted to be matched to a tracked wave path. If there was a match within the wave path confidence interval, that wave was considered to be broken (vertical lines in Figure \ref{qb_collocated}). All other waves were considered to be unbroken. $Q_b$ was then calculated following \citeA{Carini2015} as the ratio between the number of waves classified as broken ($N_{br}$) and the total number of waves ($N_t$) in the same time interval:

\begin{equation}
    Q_b = N_{br}/N_t. \label{qb_collocated_eq}
\end{equation}

In the present study, one hour of collocated video and PT data were processed using this approach. As per \citeA{Stringari2019}, data were processed in 5-minute batches. When possible, different tidal conditions were sampled to ensure a diverse dataset. This approached resulted in total number of 13,253 waves individually being classified as broken or unbroken. Some issues were observed when aligning the wave paths to waves in the PT record. Not all instances were perfectly aligned (e.g., waves 06, 23, and 26 in Figure \ref{qb_collocated}) because the alignment algorithm is intrinsically non-linear, using the dynamic time warping (DTW) method to find the best peak matches \cite{Karabiber2013,Vu2013,Hoffmann2012,Serra2014}. For some deployments, not all PTs were surveyed nor were they deployed in a perfect cross-shore orientation which also compromised the data alignment. The method to obtain $Q_b$ described above can become time-consuming because the definition of the parameters used to track the waves in the video record is not straightforward (see \citeA{Stringari2019} discussion); and, in some cases, the optimal time delay for the time-series synchronisation needed to be manually defined. The classification error using this approach compared to the manually quality-controlled dataset was of the order of 10\%, depending on the timestack quality.

\subsection{The Machine Learning Approach}\label{qb_hat}

In the previous section, a method for obtaining $Q_b$ from the collocated pressure transducer and video imagery was outlined. On one hand, this deterministic approach has the advantage of giving an exact value for $Q_b$. On the other hand, it is labour-intensive and requires manual quality control in some cases. In this section, a novel approach is developed to use PT data alone to obtain values of $Q_b$. The goal is to obtain a mapping function $g(X) = \hat{y}$ that translates an input feature vector $X$ into a predicted label $\hat{y}$ that best approximates the previously known label $y$ ($0$ unbroken waves or $1$ for broken waves). The input feature vector $X$ is represented by a series of seventeen wave-by-wave parameters that are described in Table \ref{Xvector} and Figure \ref{wbw_plot}-b. These parameters were chosen because they have been previously used in the literature to distinguish between broken and unbroken waves (e.g., \citeA{Cowell1982, svendsen2006}), or because they have historically been used to describe wave characteristics (e.g., \citeA{Ursell1953}).

\begin{table}[htp]
\caption{Wave-by-wave feature definitions (feature vector $X$). See also Figure \ref{wbw_plot} for graphical interpretations.}\label{Xvector}
\centering
\resizebox{\textwidth}{!}{%
\begin{tabular}{l l l l}
\hline
 Symbol  &  Description & Formulation & Source  \\
\hline

    $h_{cr}$ & Crest water depth & Obtained from data & \citeA{svendsen2006} \\

    $h_{tr_{a}}$ & Leading trough water depth & Obtained from data &
\citeA{svendsen2006}\\

    $h_{tr_{b}}$ & Following trough water depth & Obtained from data &
\citeA{svendsen2006}\\

    $A$ & Shape parameter A & $t({h_{cr}})-t({h_{tr_{a}}})$ &
\citeA{Cowell1982} \\

    $B$ & Shape parameter B & $t({h_{tr_b}})-t({h_{cr}})$ &
\citeA{Cowell1982} \\

    $H$ & Wave height & $H = h_{cr}-h_{tr_{a}}$ & \citeA{komar1976beach} \\

    $T$ & Wave period & $t({h_{tr_b}})-t({h_{tr_a}})$ & \citeA{komar1976beach} \\

    $\omega$ & Angular frequency & $\omega = 2\pi/T$ & \citeA{komar1976beach} \\

    $k$ & Wave number & $k=2\pi/\lambda$ &  \citeA{komar1976beach}  \\

    $\lambda$ & Wavelength & $\lambda = 2\pi/k$ & \citeA{komar1976beach} \\

    $\gamma_{tr}$ & Wave height to trough depth ratio &
$\gamma_{tr} = H/{tr}$ & \citeA{Power2010} \\

    $H/\lambda$ & Wave steepness & $H/\lambda$  & \citeA{svendsen2006} \\
    $\eta_{cr} / H$ & Vertical asymmetry &  $\eta_{cr} / H$ & \citeA{svendsen2006} \\
    $B_{0}$ & Surface shape parameter & $B_0 = \overline{(\eta/H)^2} $ &
\citeA{svendsen2006} \\

    $Ur$ & Ursell Number & $Ur = \frac{H\lambda^2}{\overline{h}}$ &
\citeA{Ursell1953} \\

    $S_k$ & Skewness - third central moment of $\eta$  & $S_k =\frac{E[(X-\mu)^3]}{\sigma^3}$ & \citeA{holthuijsen2010} \\
    $K_t$ & Kurtosis - fourth central moment of $\eta$  & $K_t =\frac{E[(X-\mu)^4]}{\sigma^4}$ & \citeA{holthuijsen2010} \\
\hline
\end{tabular}
}
\end{table}

The multi-layer perceptron (MLP) was chosen to act as the transference function $g(X)$. This class of supervised learning models has been widely used to perform image classification and natural language processing for the past three decades \cite{Haykin1994} but has only recently been applied to coastal engineering problems \cite{Zanuttigh2013, James2018}. The MLP model is organised as a series of fully-connected layers of neurons (Figure \ref{neural_net}) in which each neuron in the hidden layers represents an activation:

\begin{figure}[htp]
 \centering
 \includegraphics[width=0.95\textwidth]{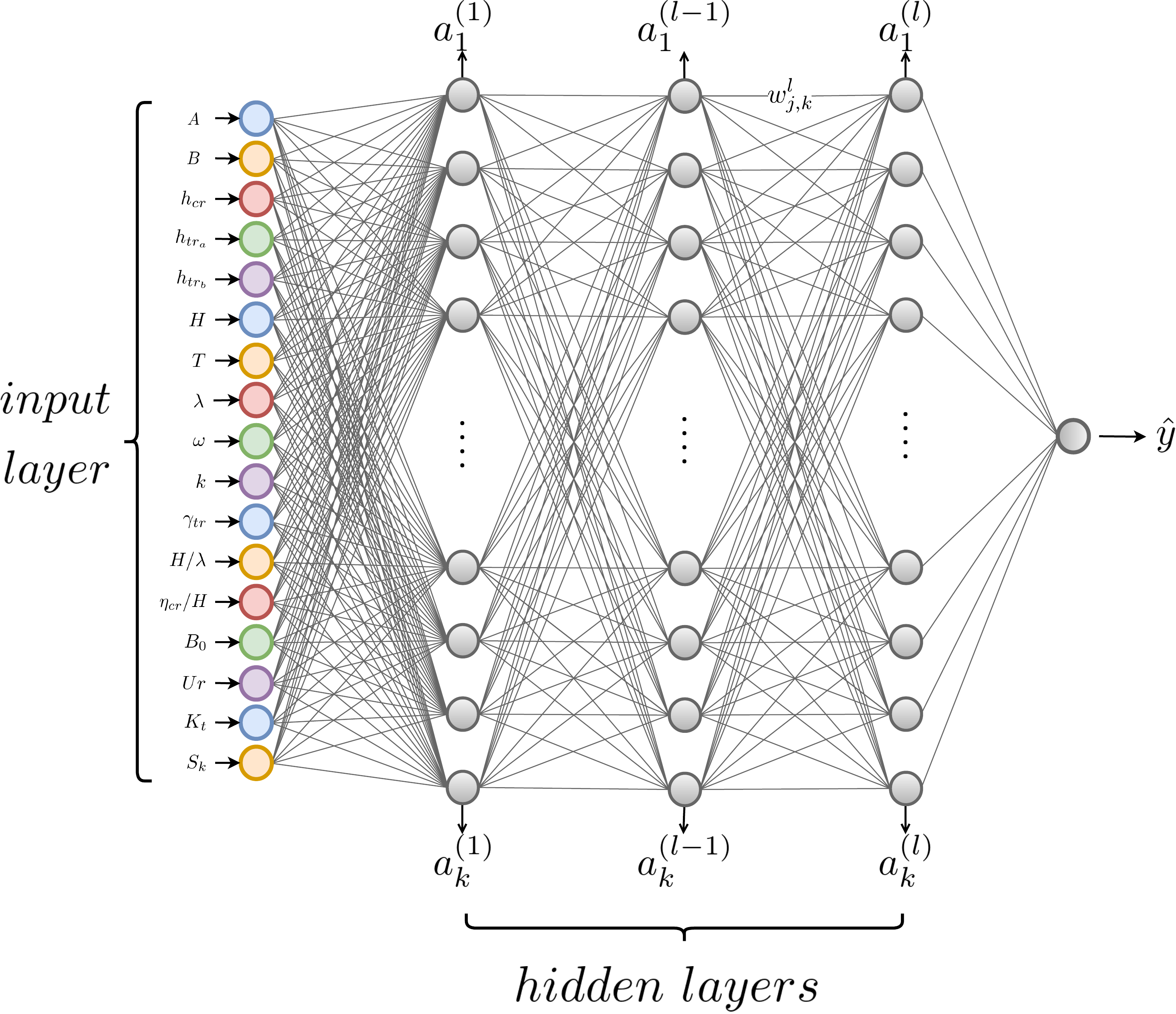}
 \caption{The multi-layer perceptron (MLP) architecture.}\label{neural_net}
\end{figure}

\begin{equation}
a^{l}_{k} = f\left( \sum_{j=1}^{n_{(l-1)}} w_{j,k}^{l} a^{l-1}_k + b^{l}_k \right)
\label{perceptron}
\end{equation}

where $a^{l}_{k}$ is the activation of a neuron $k$ in hidden layer $l$, $f$ is the ReLU (Rectified Linear Unit \cite{Nair2010}) activation function (Equation \ref{ReLU}), $w_{j,k}^{l}$ is the weight connecting neuron $k$ in hidden layer $l-1$ to neuron $k$ in hidden layer $l$, $a^{l-1}_k$ is the activation of neuron $k$ in hidden layer $l-1$, and $b^{l}_k$ is the bias added to layer $l$.

\begin{equation}
  f(z) = max(0, z). \label{ReLU}
\end{equation}

When a set of training samples $(X_1, y_1)$, $(X_2, y_2)$, $...$, $(X_n, y_n)$ is shown to the model, the MLP learns the optimal combination of weights and biases that minimises the binary cross-entropy (log-loss) cost function:

\begin{equation}
-\log p(y|\hat{y}) = -(y \log(\hat{y}) + (1 - y) \log(1 - \hat{y}))
\end{equation}

where $-\log p(y|\hat{y})$ is the negative log-likelihood of the true labels ($1$, broken wave) given the classifier's probabilistic  predictions (the activation value in the output layer). This optimisation step results in the transference function $g(X)$ that best approximates $\hat{y}$ averaged across all training samples. The learning step is accomplished via back-propagation \cite{Rumelhart1986}  using the \textit{Adam} formulation of the stochastic gradient descent (SGD) method \cite{Kingma2014}.

\subsubsection{Implementation for surf zone data}

In the present study, the training dataset consisted of the 13,253 individual waves from the unique locations described in Table \ref{expt_table} and was built using the results from the analysis describe in Section \ref{qb_quant}. Prior to the training step, each wave was visually verified to ensure that errors were not propagated into the learning algorithms (e.g., waves 6 and 23 in Figure \ref{qb_collocated}). The input vector $X$ was, therefore, a $13,253$x$17$ matrix and the $y$ output vector was a $13,253$x$1$ array. Optimal parameters for the MLP were found using a three-fold cross-validation considering 80\% of the full dataset as training samples and the remainder 20\% as testing samples. The cross-validation resulted in an optimal number of three hidden layers with 512 neurons in each hidden layer. Increasing the number of hidden layers or the number of neurons per layer resulted either in no significant performance improvement or over-fitting. Other parameters required by the \textit{Adam} algorithm were either kept unchanged from the defaults or also learnt via cross-validation. The numerical implementation was done in \textit{Keras} \cite{chollet2015keras} and has GPU (graphical processing unit) support. Because the MLP must be initialised from random weights and biases, a bootstrap procedure with 500 model runs was performed to account for the variability in the initialisation. The model run with performance closest to the median was chosen to perform further analyses (see next section for more details). This chosen pre-trained network and auxiliary programs to perform the wave-by-wave decomposition and prepare the $X$ and $y$ input arrays are freely available to the community at \url{https://github.com/caiostringari/pywavelearn/deep-learning/}.

\subsubsection{MLP Verification and Validation}\label{mlp_valid}

Different approaches were used to verify and validate the MLP results. In all subsequent analyses, only the test dataset is considered. The first approach used the $F_1$ score \cite{Rijsbergen} to evaluate each bootstrap run. This metric should be more robust than the standard classification score because it takes into account both precision (the number of correct positive results divided by the number of all positive results returned by the classifier) and the recall (the number of correct positive results divided by the number of all relevant samples).  The median $F_1$ score was 86.79\% for broken waves and 86.97\% for unbroken waves (higher $F_1$ scores indicate better performing models).

\begin{figure}[htp]
  \centering
  \includegraphics[width=0.95\textwidth]{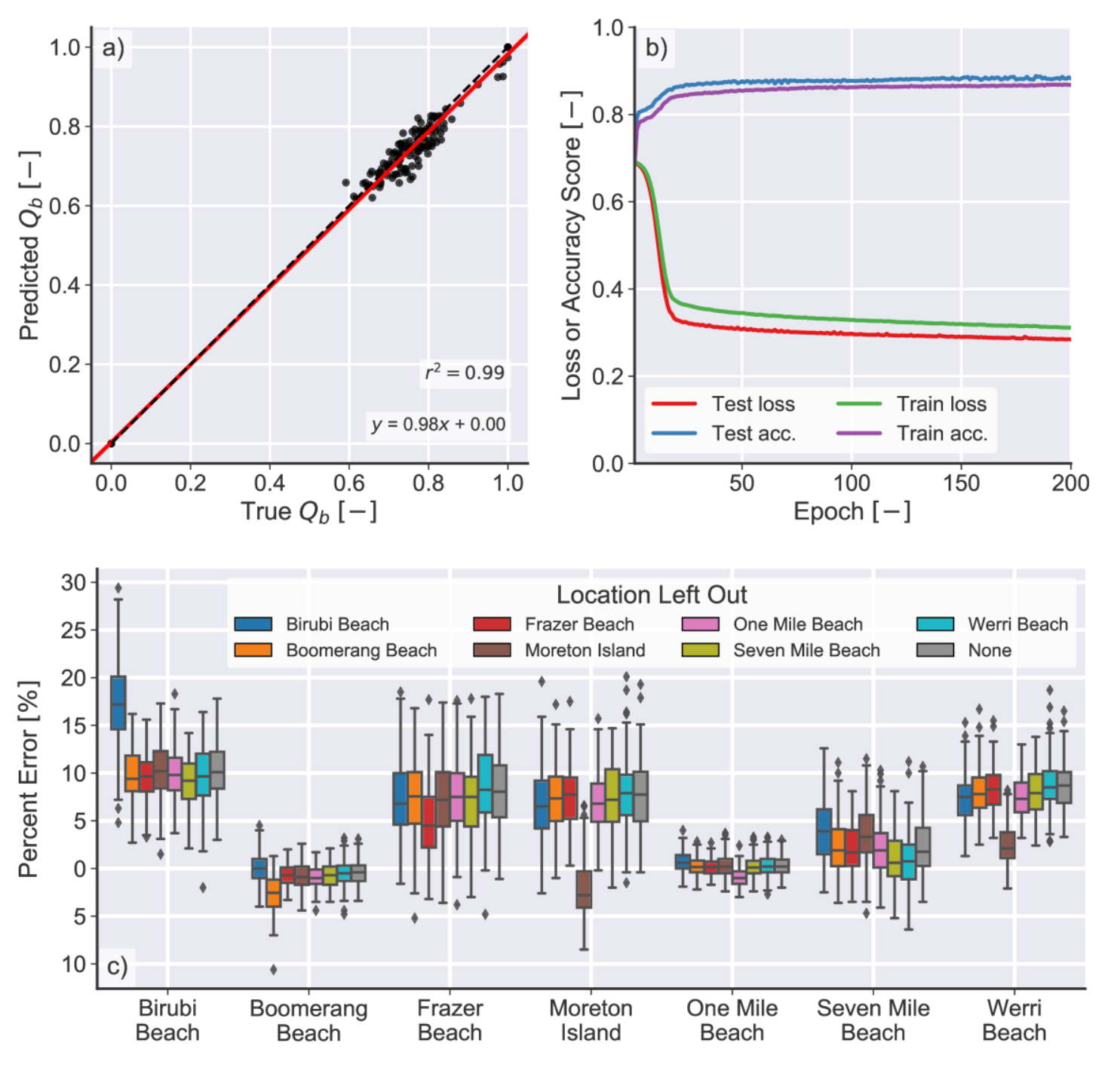}
  \caption{a) Comparison between the measured and  predicted $Q_b$ for the test dataset only computed using the median-performing MLP. The red line shows the linear regression analysis, the red swath shows the 95\% confidence interval, and the black dashed line shows the one-to-one correspondence line. b) Accuracy scores and log-loss function values for the first 200 epochs of training for the median-performing MLP. c) Box-and-whisker plot of the mean percentage error (MPE) for the 500 leave-one-out bootstrap runs sorted by location. The whiskers represent one standard deviation interval and the diamond markers, outliers. The MPE was calculated using only the test dataset, i.e., the left-out location.}\label{qb_bootstrap}
\end{figure}

To provide further evidence that the MLP is a valid surrogate for the direct method, a comparison between $Q_b$ calculated from both methods was conducted (Figure \ref{qb_bootstrap}-a). The results of this analysis showed a solid correspondence between both methods with $r^2=0.99$ and $p<<0.001$. To ensure that was no significant over-fitting, the training history was also recorded (Figure \ref{conclusion}-b). These results strongly indicated that there was no over-fitting, and that the model generalised well for the test data. Note that even after 200 training epochs, the MLP was still slowly learning. Given these promising results, the median-performing MLP was used to classify the 333,732 individual waves measured in all experiments described in Table \ref{expt_table}, totalling 3,639 unique 15-minute timeseries (or data runs) within all PTs . The results from this classification will be used in the analyses starting from Section \ref{qb_results}.

To test the effect of inter-beach differences in the learning step, a second, independent, bootstrap procedure was conducted. For each run, one of the locations was left out of the training dataset, the MLP was trained, and then used to classify the data from the left-out location. The mean percentage error (MPE) between $Q_b$ calculated using the true and the predicted wave labels was used as the evaluation metric (Figure \ref{qb_bootstrap}-c). The highest errors were seen when Birubi Beach data was left out of the training step and data from the same location was classified. This was an indication that the model did not generalise as well for this location. For all other beaches, the averaged MPE was below 10\% which is a good indicator that the MLP generalised well. Some counter-intuitive results were also observed, e.g., when Moreton Island was left out and data from same location was classified, the classification was better. Similar trends occurred for Boomerang Beach and Werri Beach. See Section \ref{qb_bootstrap} for a discussion of the possible causes for these results. Nonetheless, these results strongly indicate that the MLP can be used as a valid substitute for the collocated PT-video method presented in Section \ref{qb_quant}, with the clear advantage of being able to classify hundreds of thousands of waves in virtually no computational time.

\subsubsection{Analysis of the Neural Network Structure}\label{neural_net_struct}

Neural networks are usually seen as black-boxes in which only node(s) of the output layer bear meaning to the classification problem in question. However, there has been a recent increase in research on how to detect which features are most important to the network and how these features relate to the final classification results (e.g., Google's deep dream experiment \cite{Simonyan2013}). Unfortunately, there is still no consensus on which techniques should be used to perform such a task. One way to tackle this problem is to use the concept of feature importance from tree-based models (e.g., decision trees) adapted to neural networks. In this study, the feature importance of the variables in the input layer was obtained by zeroing the contribution of each variable in the input vector $X$, re-training the network, and classifying data in the test dataset. The feature importance measure was then obtained as the difference between the score from the median run (see Section \ref{mlp_valid}) and the score obtained when each variable was zeroed. Due to the random initialisation of the MLP, this procedure was repeated 500 times for each variable to obtain statistically significant results.

\begin{figure}[htp]
  \centering
  \includegraphics[width=0.95\textwidth]{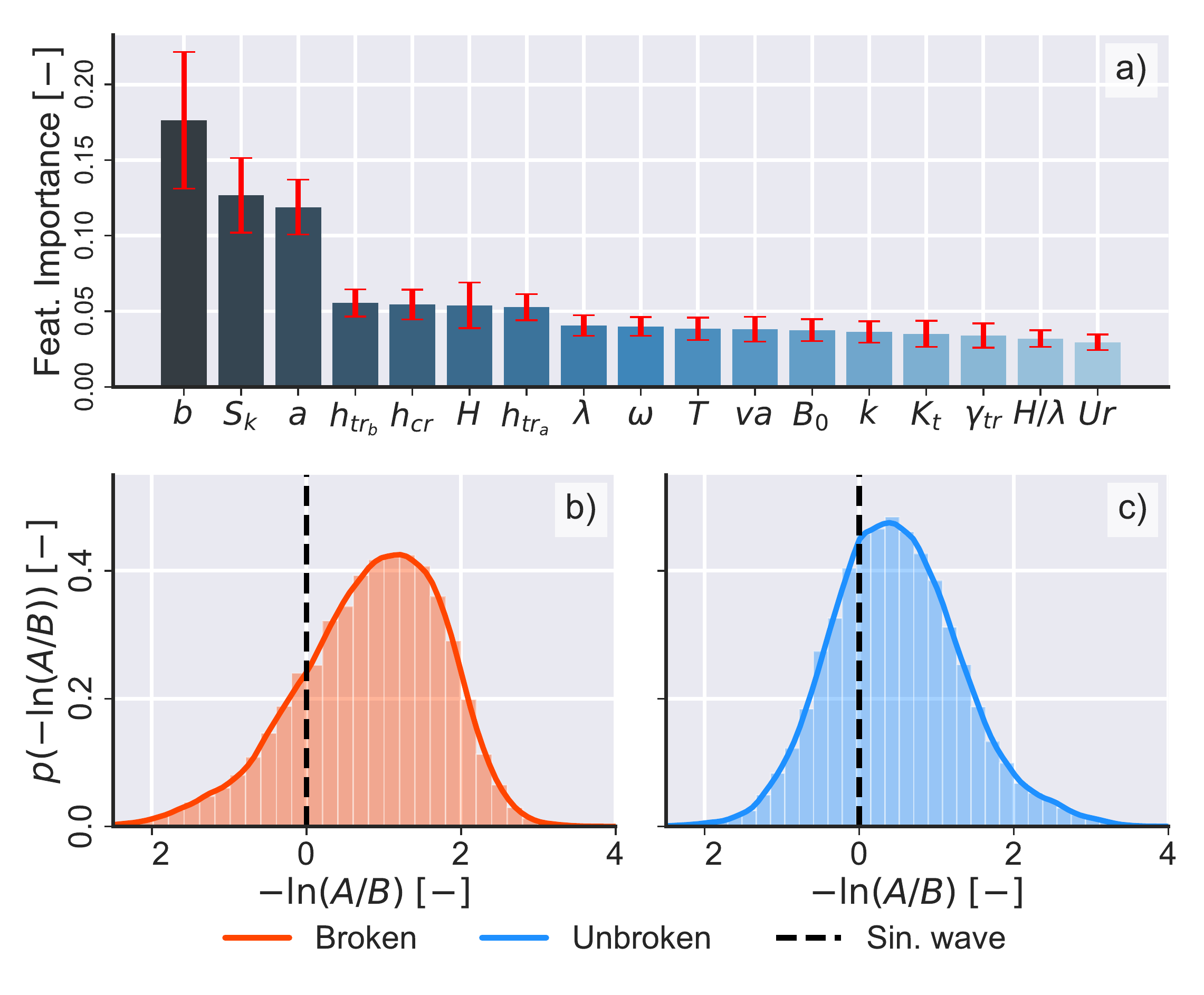}
  \caption{a) Averaged feature importance across the 500 runs (bars) and standard deviation for each feature in the MLP input layer (red error bars). b) $p(-\ln(A/B))$ for broken waves only. c) $p(-\ln(A/B))$ for unbroken waves only. The black dashed line in b) and c) indicates $-ln(A/B)$=$0$, the expected value for a sinusoidal wave.}\label{feature_imporances}
\end{figure}

The most important features were the shape parameters $A$ and $B$ \cite{Cowell1982} and the skewness of the surface elevation. These parameters combined accounted for 45\% of the total feature importance (Figure \ref{feature_imporances}-a) thus demonstrating that the MLP was identifying that the differences in  wave shape were the most important feature when classifying waves into broken or unbroken. This is a sensible result as broken waves in the surf zone usually present a characteristically skewed, triangular shape (saw-tooth shape), whereas unbroken waves are more symmetrical and less skewed \cite{Svendsen1978, Cowell1982}.

Analysis of the PDF of the combined shape parameter  $-\ln(A/B)$ \cite{Cowell1982} showed that the majority of broken waves had shapes similar to ``bore-like'' waves (Figure \ref{feature_imporances}-b). Conversely, the peak of the distribution of unbroken wave shapes was close to the expected value for a sinusoidal wave (Figure \ref{feature_imporances}-c), thus being mainly composed by less skewed waves. Furthermore, the two-sample Kolmogorov-Smirnov (K-S) test indicated that the PDFs shown in Figure \ref{feature_imporances}-a) and \ref{feature_imporances}-b) are statistically significantly different with $p<<0.001$. This is another good indication that the MLP is indeed capable of separating broken and unbroken waves based on a plausible combination of shape parameters and that it is not learning based on an arbitrary combination of weights and biases found by the SGD algorithm.

\section{Results}\label{results}

\subsection{Results of field observations}\label{qb_results}

\begin{figure}[htp]
  \centering 
  \includegraphics[width=0.99\textwidth]{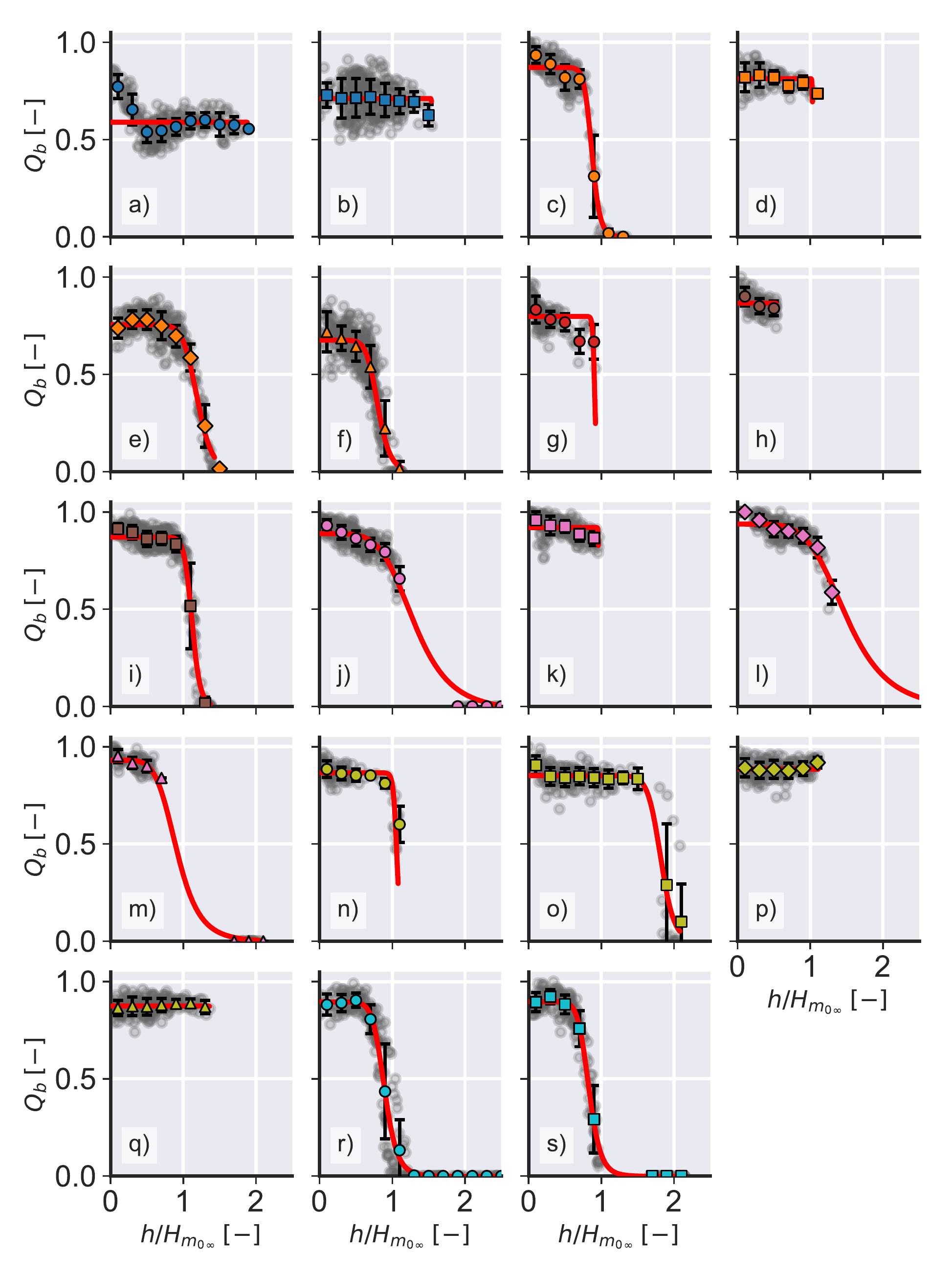}
  \caption{(Caption next page.)}\label{qb_by_depth}
\end{figure}

Figure \ref{qb_by_depth} shows the variation of $Q_b$ relative to averaged water depth normalised by offshore wave height, i.e., relative water depth, which will be used as a proxy for the cross-shore distance hereafter. It should be noted, however, that this approach may not be wholly appropriate for barred beach profiles because depths on either side of the bar would appear in the same location on the $x$-axis. The most notable feature observed was that, for similar relative water depths, there was great variability in $Q_b$ at all locations analysed (grey markers in Figure \ref{qb_by_depth}). This observation of the behaviour of $Q_b$ is novel in the literature but is consistent with observations of other natural surf zone parameters \cite{Power2010, Postacchini2014, Martins2017a} and with the chaotic behaviour of breaking waves \cite{Wei2018}. $Q_b$ values were binned at 0.1 relative depth intervals (coloured markers in Figure \ref{qb_by_depth}) and a four-parameter logistic curve \cite{Richards1959} was fitted to the data (thick red lines in the panels in Figure \ref{qb_by_depth}) to aid with visualisation of the general cross-shore structure of $Q_b$. For the beaches where the whole surf zone was sampled, the logistic fit agreed well with the observations (i.e., at Boomerang Beach, Moreton Island, Werri Beach, and One Mile Beach) but this fitting method could be inappropriate to model $Q_b$ when the surf zone was only partially sampled (i.e., at Frazer Beach and Birubi Beach, and Seven Mile Beach).

An inverse linear trend was observed in the inner surf zone in some cases (e.g., Figure \ref{qb_by_depth}-p) and Figure \ref{qb_by_depth}-q). These trend are, however, not statistically significantly different from an averaged constant value of $Q_b$. Interestingly, terminal values of $Q_b$ were often observed to be significantly less than one, which is consistent with a $p(H)$ in which small unbroken waves reach the surf-swash boundary or could be caused by shoreline reflection \cite{Martins2017}. Such terminal values would not be predicted by the models presented in Section \ref{QB_Theory_01} due to the constraint $\lim_{h \rightarrow 0} Q_b(h) = 1$. Finally, the outer limit of the surf zone was observed to be in the range  $1<h/H_{{m_0}_{\infty}}<2$ instead of the value of $2<h/H_{{m_0}_{\infty}}<3$  suggested by several other publications \cite{Thornton1982, Thornton1983, Power2010, Ruessink1998}.

\addtocounter{figure}{-1}
\begin{figure}[t!]
  \caption{Variation of $Q_b$ by averaged water depth normalised by the offshore wave height (relative water depth). a) and b) Birubi Beach (05/07/2017 and 06/07/2017). c), d), e) and f) Boomerang Beach (19/09/2016 to 22/09/2016). g) Frazer Beach (24/04/2018), h), and i) Moreton Island (19/12/2016 and 20/12/2016). j), k), l) and m) One Mile Beach (04/08/2014 to 07/08/2014). n), o), p) and q) Seven Mile Beach (13/08/2014, 14/08/2014, 19/08/2014, and, 20/08/2014,). r) and s) Werri Beach (15/08/2014 and 16/08/2014). The coloured markers show binned values at 0.1 relative depth intervals, the error bars show the standard deviation at each bin, the gray circles show the calculated $Q_b$ values, and the red continuous lines show the four-parameter logistic fit to the data.}
\end{figure}

\subsection{Investigation of the inter and intra-beach Variability of $Q_b$}\label{env_fact}

The variability seen in Figure \ref{qb_by_depth} is examined bellow with respect to three surf zone parameters: tidal cycle, infragravity wave energy, and beach morphology. These three parameters were chosen because, given an appropriated predictive model (in this case, a decision-tree model - not shown), they explained 90\% of the variability seen in the cross-shore variation of $Q_b$. Similar trends were observed for all locations in the dataset but, for brevity, only the most representative cases for each parameter are analysed in detail below.

\subsubsection{Tidal cycle}\label{tides}

\begin{figure}[htp]
  \centering
  \includegraphics[width=0.95\textwidth]{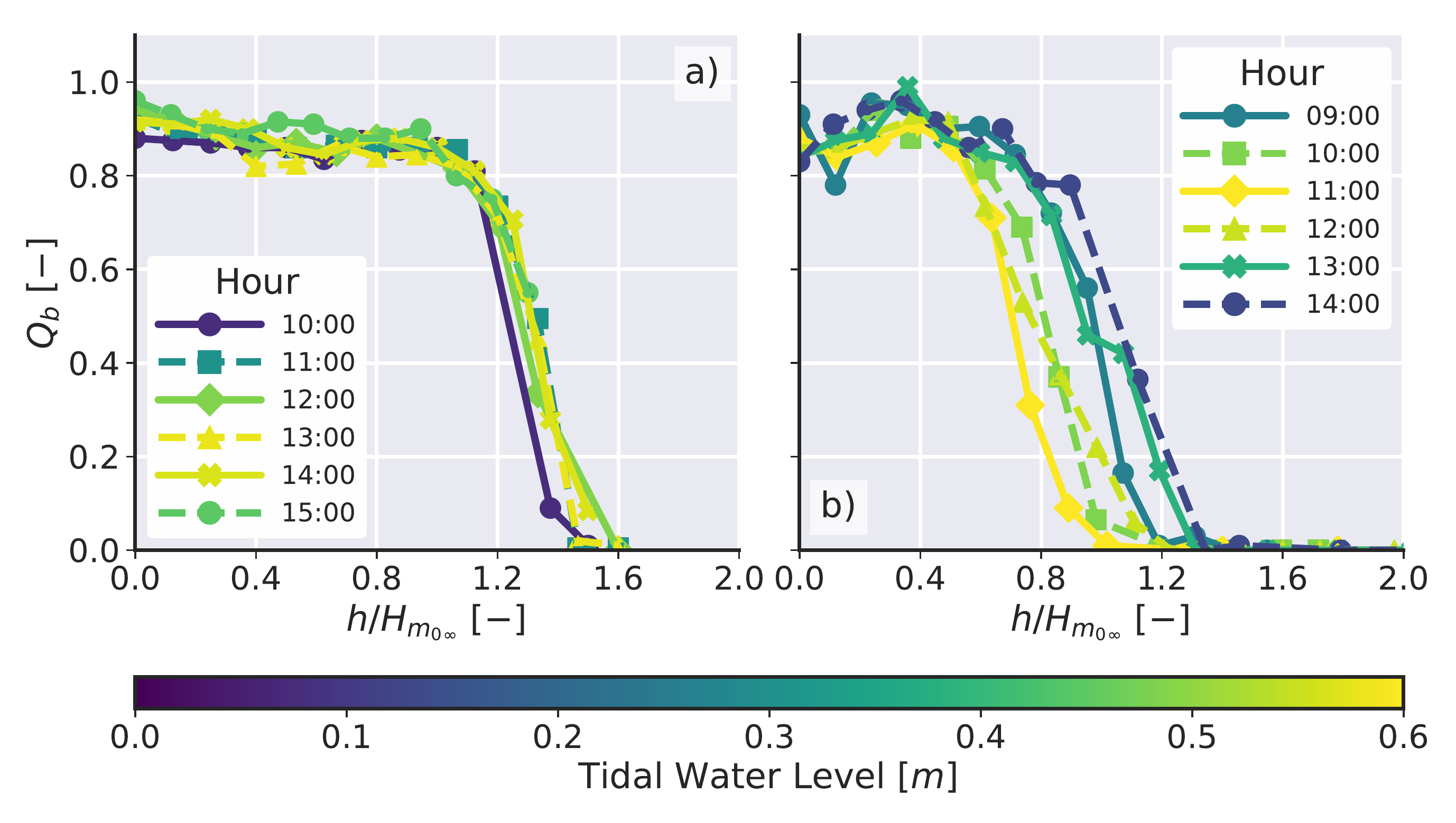}
  \caption{Variation of the fraction of broken waves by relative water depth, hour of the day, and tidal water level at a) Moreton Island (20/12/2016), and b) Werri Beach (16/08/2014).}\label{tidal_influence}
\end{figure}

The tidal influence on $Q_b$ was analysed using data from Moreton Island (20/12/2016) and Werri Beach (16/08/2014). These two deployments were chosen because: 1) they had a large number of PTs in the surf zone during the tidal cycle, 2) they had contrasting beach profile characteristics (see Figure \ref{profiles}), and 3) they had offshore wave conditions that remained approximately constant during the duration of the deployment. For each 15-minute run, data were extracted from the MLP classified dataset, binned into 0.1 water depth intervals, grouped by hour, and coloured by tidal water level obtained using the Pacific Ocean solution of Oregon State University's Tidal Prediction Software (OTPS) \cite{Egbert2002}. The vertical datum used for tides was the same as in OTPS. The results of this analysis are shown in Figure \ref{tidal_influence}.

For the gently sloping beach (Moreton Island), no influence of tidal control on $Q_b$ was observed (Figure \ref{tidal_influence}-a). For this profile, tides may be only responsible modifying $Q_b$'s behaviour in deeper water depths, closer to the seaward end of the surf zone. As such, lower values of $Q_b$ were systematically observed during low tide. In contrast, there was a clear tidal signature in the cross-shore evolution of $Q_b$ for the steep profile (Werri Beach, Figure \ref{tidal_influence}-b). Where, during low tide, the main break shifted towards the terrace portion of the beach profile (see panel g) in Figure 3) which caused the $Q_b$ curves to be shifted toward deeper relative depths. At higher tidal water levels, the main break point shifted shoreward to a very steep portion of the profile, causing the $Q_b$ curves to be shifted toward shallower water depth, and therefore causing the surf zone to become narrower. On the steep profile case, the tidal cycle drove variations in $Q_b$ of up to 70\% at similar relative water depths (e.g., at $h/H_{{m_0}_{\infty}}=0.9$). A possible explanatory mechanism for this dynamics is discussed in Section \ref{diss_qb}

\subsubsection{Infragravity energy}\label{ig_energey}

\begin{figure}[htp]
  \centering
  \includegraphics[width=0.95\textwidth]{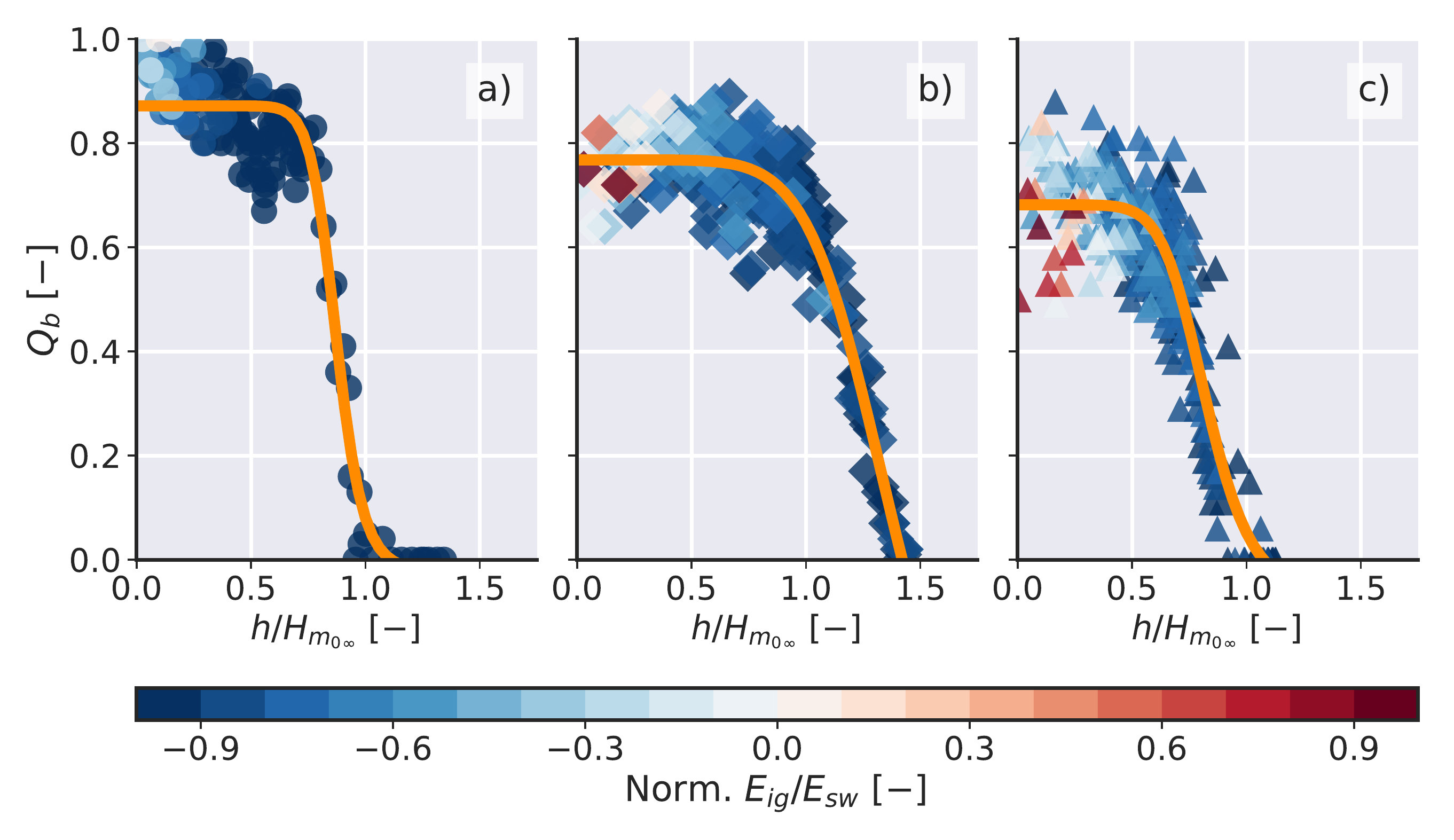}
  \caption{Variation of the $Q_b$ by relative water depth and coloured by the ratio between infragravity and sea-swell spectral energy ($E_{ig}/E_{sw}$) normalised by the total energy ($E_{ig}$+$E_{sw}$) for each deployment. The thick orange line shows the logistic fit to the data. a) 19/09/2016, b) 21/09/2016, and c) 22/09/2016. The 20/09/2016 experiment was excluded because the surf zone was not fully sampled.}\label{qb_by_depth_and_ig}
\end{figure}

Three deployments at Boomerang Beach were chosen to highlight the influence of infragravity waves on $Q_b$. These deployments were chosen because they presented comparable offshore wave conditions, beach profiles, and tidal water levels but three different trends in the cross-shore variation of $Q_b$. Similar but less pronounced trends were observed in all other locations (not shown). For the analysed data (Figure \ref{qb_by_depth_and_ig}), the ratio between the spectral energy in the infragravity ($E_{ig}$) and sea-swell ($E_{sw}$) frequency bands was calculated as

\begin{equation}
    \frac{E_{ig}}{E_{sw}} = \frac{\int^{f=0.04}_{f=0.004}  E(f)  df} {\int^{f=0.04}_{f=1}  E(f)  df}
\end{equation}

in which $E$ is the total energy in each frequency band for individual 15-minutes timeseries. The results of this analysis are shown in Figure \ref{qb_by_depth_and_ig}.

In general, $E_{ig}$ increased towards shallower water depths for all data runs, which is consistent with several previous observations (see \citeA{Bertin2018a} for a recent review). During the first deployment (19/09/2016), the surf zone was dominated by waves in the sea-swell frequency and the terminal value of the fitted $Q_b$ curve was close to 1 (Figure \ref{qb_by_depth_and_ig}-a, $E_{ig}/E_{sw} \leq 0$). Subsequently, infragravity wave energy started to dominate in shallower water depths, causing the terminal value of the $Q_b$ curve to lower (Figure \ref{qb_by_depth_and_ig}-b, $E_{ig}/E_{sw} \geq 0$; 21/09/2016). Finally, the inner surf zone was strongly dominated by infragravity waves during the last deployment, which caused the terminal values of the $Q_b$ curves to reach it lowest (Figure \ref{qb_by_depth_and_ig}-c, $E_{ig}/E_{sw}>0.5$; 22/09/2016). These results suggest that increasing  $E_{ig}$ may lower the terminal values of $Q_b$. Possible causes for the infragravity control on $Qb$ are discussed in Section \ref{diss_qb}.

\subsubsection{Beach morphodynamics}\label{res_omega}

\begin{figure}[htp]
  \centering
  \includegraphics[width=0.95\textwidth]{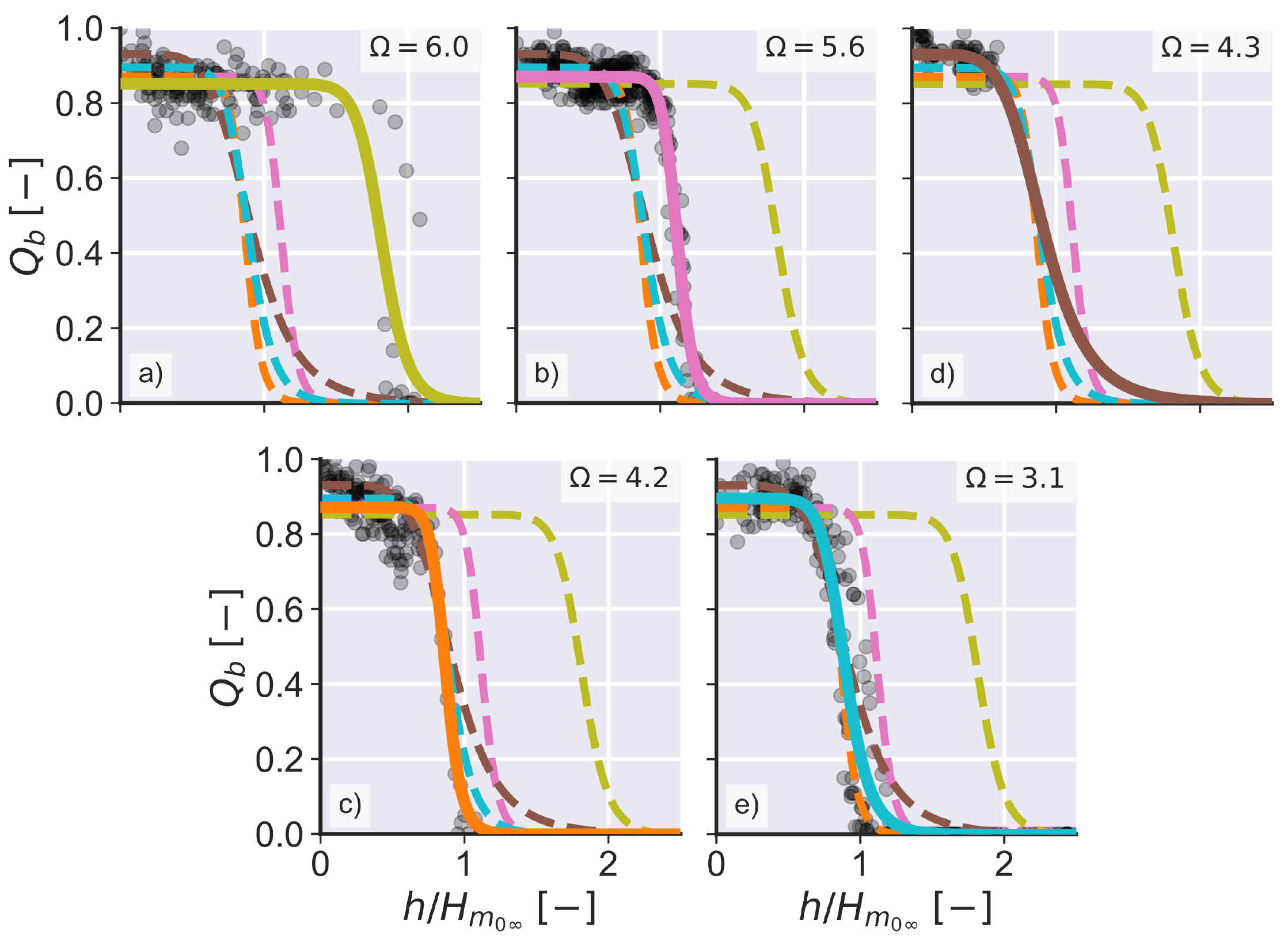}
  \caption{The link between $Qb$ and $\Omega$. The panels are sorted by decreasing $\Omega$ (dissipative $\rightarrow$ intermediate). a) Seven Mile Beach (14/08/2014), b) One Mile Beach (07/08/2014), c) Moreton Island (20/12/2016), d) Boomerang Beach (19/09/2016), and e) Werri Beach (16/08/2014). In each panel the thick coloured line shows the fitted $Q_b$ curve at the analysed location, the dashed coloured lines show the fitted $Q_b$ curves for the other locations, and the black markers show the calculated $Q_b$ values.}\label{qb_by_depth_and_omega}
\end{figure}

The link between $Q_b$ and \citeA{Wright1984} beach morphodynamic model was investigated using the parameter $\Omega$, defined as :

\begin{equation}
   \Omega =  \frac{H_{m{_0}}}{W_s T_{m{_{01}}}}
\end{equation}

in which $H_{{m_0}}$ is the surf zone significant wave height,  $T_{m{_{01}}}$ is the surf zone significant wave period, and $W_s$ is the sediment fall velocity \cite{Dean1973, Gourlay1968}. Data from one day at the locations where the surf zone was fully sampled were used in this analysis. The fitted $Q_b$ curves grouped into two main clusters: one for dissipative beaches ($\Omega>=6$, Seven Mile Beach; Figure \ref{qb_by_depth_and_omega}-a), and another for intermediate ($2<\Omega<6$, Boomerang Beach, One Mile Beach, Werri Beach, and Moreton Island, Figure \ref{qb_by_depth_and_omega}-b to c) which is in very good agreement with \citeA{Wright1984} model. These results also correlated well with the observed breaker types and surf zone widths from the video data. At Seven Mile Beach, the surf zone was observed to be wide and the waves had time to develop into fully formed bores which dissipate most of incoming energy. For the other locations, the surf zone was observed to be narrower and dominated by plunging breakers, which were not observed to develop into fully developed bores.

\subsection{Broken Wave Height PDF}\label{Hbrk_results}

\begin{figure}[htp]
  \centering
  \includegraphics[width=0.95\textwidth]{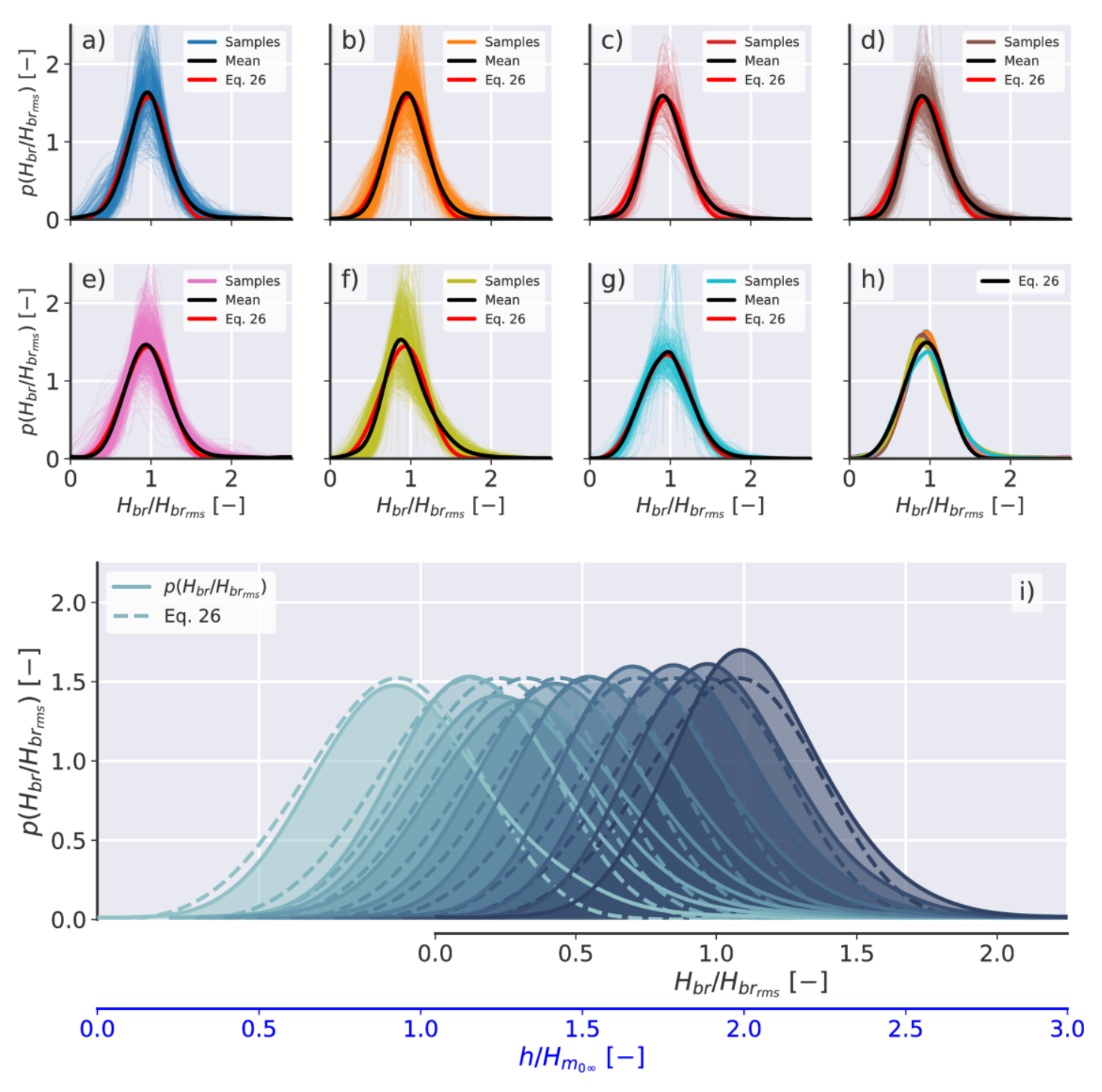}
  \caption{Approximated PDF of broken waves ($p(H_{br})$) via KDEs for a) Birubi Beach, b) Boomerang Beach, c) Frazer beach, d) Moreton Island, e) One Mile Beach, f) Seven Mile Beach, and g) Werri Beach. The coloured lines in each pannel show $p(H_{br})$ calculated by a Gaussian KDE for each data run. The thick continuous black lines show the mean $p(H_{br})$ at each location, and the continuous red lines show Equation \ref{weibull_data}. h) Averaged PDF for all locations (coloured lines) and Equation \ref{weibull_data}  (black line). i) PDF of broken waves for all locations grouped by relative water depth deciles from shallower (lighter shades) to deeper (darker shades) deciles. The solid lines indicate the averaged PDF at each decile and the dashed lines the approximation from Equation \ref{weibull_data}. Each PDF was plotted centred at the respective relative water depth decile (blue axis). The secondary (black) axis in i) is the same x-axis as in the a) to h) and repeats for each PDF.}\label{pH_broken_waves}
\end{figure}

Here, the data obtained from the previous analysis are used to develop a novel definition for the PDF of broken waves. For each location and each data run, the wave heights of all the broken waves ($H_{br}$) were extracted, and $p(H_{br})$ was approximated via non-parametric kernel density estimations (KDE) \cite{Kim2012} (Figure \ref{pH_broken_waves}-a to g). Variability between PDFs at a location was observed (coloured lines in Figure \ref{pH_broken_waves}), consistent to the results presented in Section \ref{qb_results}.  However, when the mean PDF for each location was calculated, the results were remarkably consistent across all locations (Figure \ref{pH_broken_waves}-h). On average, $p(H_{br})$ could be approximated by a Weibull PDF with $\kappa$ $\approx$ $4$ and scale $H_{br}/H_{br_{rms}}$. These optimal values were calculated as the average of the shape and scale parameters found after fitting Equation \ref{pH_B98_W} to $p(H_{br})$ approximated via the KDEs. From these results, it is possible to obtain an analytic expression for $p(H_{br})$:

\begin{equation}
  p\left( \frac{H_{br}}{H_{br_{rms}}} \right) = 4\left( \frac{H_{br}}{H_{br_{rms}}} \right)^3 \exp{\left[-\left( \frac{H_{br}}{H_{br_{rms}}} \right)^4 \right]}. \label{weibull_data}
\end{equation}

\noindent Note that $p(H_{br})$ is not equivalent to $p_b(H)$ introduced in Section \ref{QB_Theory_01} because $p_b(H)$ is not a true PDF whereas $p(H_{br})$  is.

When Equation \ref{weibull_data} was compared to averaged PDFs (black lines in Figure \ref{pH_broken_waves}) using the K-S test, there was no significant statistical difference between them at the 95\% confidence interval. Moreover, when the t-test for the mean was applied to compared individual PDFs (coloured lines in Figure \ref{pH_broken_waves}) to Equation \ref{weibull_data}, it was found that in 96.4\% of the cases there was no statistical difference between then at the 95\% confidence interval. To further test the robustness of Equation \ref{weibull_data}, it was also compared to the mean PDF at each relative depth decile considering data from all locations (Figure \ref{pH_broken_waves}-i). Except for the deepest decile, the approximations via Equation \ref{weibull_data} were not statically different to the averaged PDF at the 95\% confidence interval. This is an important finding because from Equation \ref{weibull_data} the $p(H)$ can be analytically transformed into $p(H_{br})$.

Let $f(x)$ and $g(y)$ be two functions that satisfy the following properties: $f(x)$ and $g(y)$ must be single valued for all $x$ and $y$, $f(x) \geq 0$ for all $x$ and $g(y) \geq 0$ for all $y$, and $\displaystyle \int_{0}^{\infty}f(x)dx = 1$ and $\displaystyle \int_{0}^{\infty}g(y)dy = 1$. If these properties hold, the following is valid:

\begin{equation}
g(y) = f(x(y)) \left| \frac{dx}{dy} \right| \label{transference}
\end{equation}

in which $f(x(y))$ is a transference function. Substituting $f(x)$ = $p_1 \left( \frac{H}{H_{rms}} \right)$ and $g(y)$ = $p_2 \left( \frac{H_{br}}{H_{br_{rms}}} \right)$ results in:

\begin{equation}
    p_2 \left( \frac{H_{br}}{H_{br_{rms}}} \right) = p_1 \left(\frac{H}{H_{rms}} \left( \frac{H_{br}}{H_{br_{rms}}} \right) \right) \left| \frac{d \left( \frac{H}{H_{rms}} \right) }{d \left( \frac{H_{br}}{H_{br_{rms}}} \right) } \right|. \label{true_model}
\end{equation}

This result is key because it allows for a mathematically rigorous transformation of the PDF of all waves into the PDF of broken waves. Solving this differential equation for the exact transformation function and applying it directly into Equation \ref{E} is, however, beyond the scope of this paper and will be attempted in a follow-up publication.
 
% \footnote{Using the integral substitution method describe in \href{https://en.wikibooks.org/wiki/Probability/Transformation_of_Probability_Densities}{here}.}

\subsection{Assessing existing $Q_b$ Models}\label{model_results}

\begin{figure}[htp]
  \centering
  \includegraphics[width=0.95\textwidth]{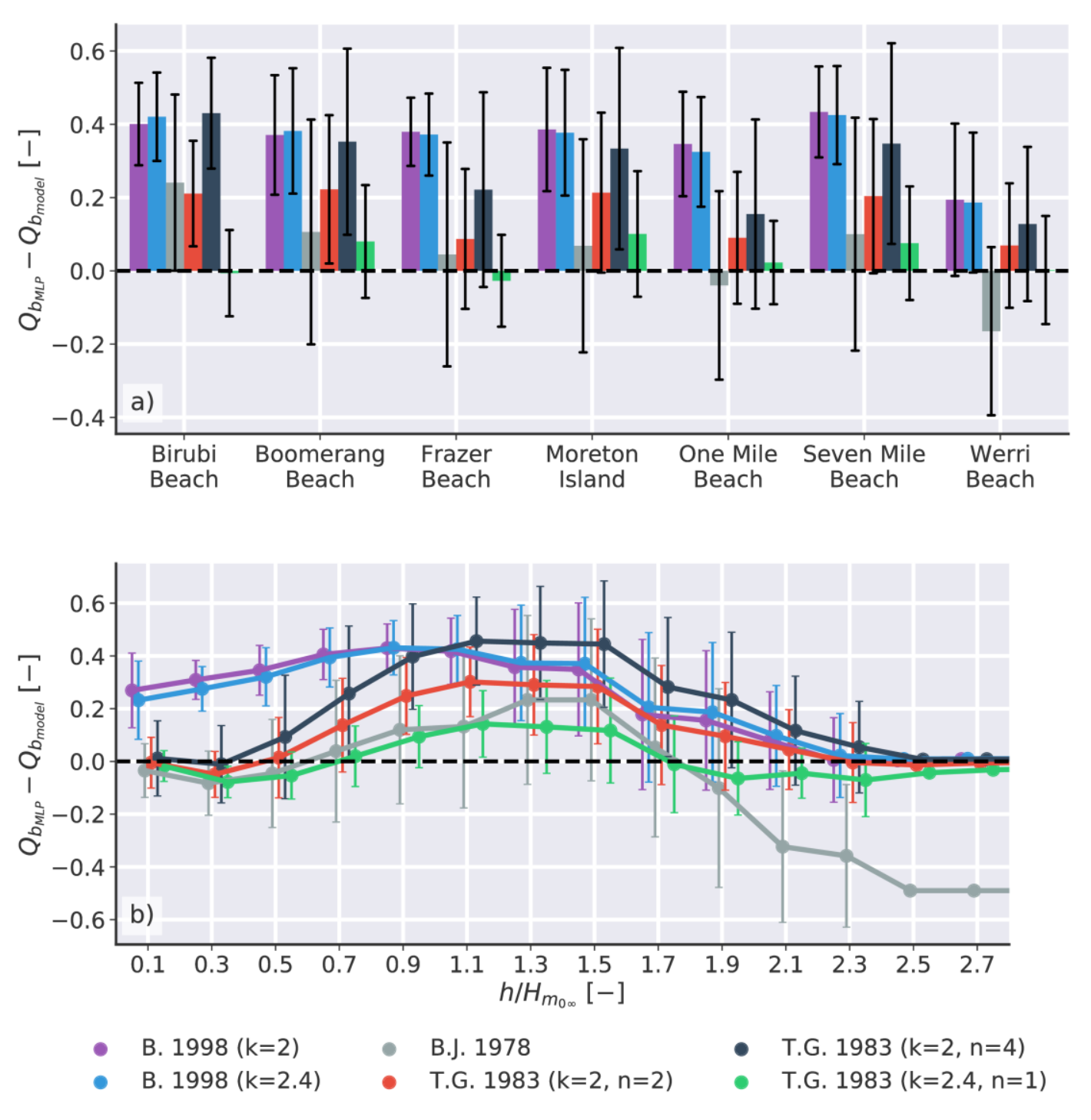}
  \caption{a) Averaged residual between the measured and theoretically predicted $Q_b$ grouped by location (bar heights) and respective standard deviation (error bars). b) Averaged residuals at 0.2 relative depth intervals (data for all locations). The vertical error bars indicate the standard deviation. The points for each model in b) were slightly shifted around the bin average to aid visualisation.}\label{qb_residual}
\end{figure}

The models presented in Section \ref{QB_Theory_01} were assessed using $Q_b$ calculated from the MLP classified data (see Section \ref{mlp_valid}). For each data run, the residuals between the calculated ($Q_{b_{MLP}}$) and the theoretically predicted ($Q_{b_{model}}$) were obtained and the results are shown in Figure \ref{qb_residual}-a). As anticipated by the results seen in Section \ref{qb_results}, all models performed poorly because they cannot account for intra-depth variability given similar offshore conditions nor terminal values of $Q_b\neq1$. The models with the lowest averaged residuals were the original (with $k=2$ and $n=2$) and the modified TG83 models (with $k=2.4$ and $n=1$). Both original and modified B98 models performed poorly when considering only the averaged residuals but, despite this, were the most consistent models, showing no clear bias toward one particular beach. On the contrary, BJ78 presented the highest inter- and intra-location variability in the residuals. Because the models differ in the number of input parameters, the averaged residuals alone may not be an appropriate comparative metric. Using the Akaike information criterion (AIC; \citeA{Akaike1974,Aho2014}) to account for the different number of input parameters, it was confirmed that the best performing model was TG83 with $\kappa$=2.4 and $n=1$. Table \ref{model_res} shows the full analysis.

\begin{table}[htp]
 \centering
 \caption{Model assessment using the AIC  as metric. ($AIC_{min}$ is the minimum AIC).}\label{model_res}
 \begin{tabular}{llcccc}
  \toprule
   & model & Parameters & N. par.  & $AIC$ & $AIC-AIC_{min}$ \\
  \midrule
   1 & TG83 ($\kappa$=2.4, n=1) & $\gamma$, $\kappa$, $n$ & 3 & -13590 &     0 \\
   2 & TG83 ($\kappa$=2, n=2)   & $\gamma$, $n$ & 2 &  -9788 &  3801 \\
   3 & BJ78              & $H_{max}$ & 1 &  -8593 &  4996 \\
   4 & TG83 ($\kappa$=2, n=4)   & $\gamma$, $n$ & 2 &  -6797 &  6792 \\
   5 & B98 ($\kappa$=2.4)       & $\gamma$, $\kappa$ & 2 &  -6708 &  6882 \\
   6 & B98 ($\kappa$=2)         & $\gamma$ & 1 &  -6696 &  6894 \\
  \bottomrule
\end{tabular}
\end{table}

Residuals were also analysed by relative water depth (Figure \ref{qb_residual}-b). The data for all locations were grouped, binned into 0.2 relative depth intervals, and the residuals were calculated as before. The results from this analysis showed that BJ78 greatly over-predicted $Q_b$ at deeper water depths, whereas both B98 model parametrisations under-predicted $Q_b$ at shallower water depths. For all models, the greatest averaged residuals (circular markers in Figure \ref{qb_residual}-b) were observed in the mid-surf zone (relative water depth range of 0.7-1.9). B98's systematic underestimation of $Q_b$ at shallower water depths could be yet another factor that affected the total energy dissipation in this model in addition to simplification of $H^3/h \approxeq H^2$ as reported by \citeA{Alsina2007} and \citeA{Janssen2007}. No correlations were found between the errors seen in the models and other surf zone parameters (not shown).

\section{Discussion}\label{discussion}

In this paper, a comprehensive review of the theoretical formulations for the fraction of broken waves ($Q_b$) was initially presented (Section \ref{QB_Theory_01}). It was evident from this that little testing of these formulations against natural surf zone data had been performed and further testing was required. Data from seven beaches across 19 deployments were collected and used to address this shortcoming (Section \ref{data_collection}). Two novel methods were developed to obtain $Q_b$. Firstly by using collocated PT and video data and, secondly, by using robust machine learning techniques (Sections \ref{qb_quant} and \ref{qb_hat}). Using the second approach, 333,732 waves were classified which resulted in 3639 unique 15-minute timeseries (data runs) being analysed. Based on this dataset, the cross-shore structure of $Q_b$ was analysed to access the influence of environmental parameters on $Q_b$ and develop novel definitions of the PDF for broken waves. Finally, three widely used theoretical models were assessed. In the next sections, the methods used throughout this paper and the results they produced are discussed.

\subsection{Cross-shore variability of $Q_b$ and Environmental Forcing}\label{diss_qb}

The analysis of the cross-shore structure of $Q_b$'s showed great intra-depth variability (Figure \ref{qb_by_depth}) which seems not to have been previously observed in the literature. To the author’s knowledge, only TG83 measured $Q_b$ on natural surf zones using in-situ data, and, in their work, only four cross-shore observations from one beach were shown (their Figure 11). Such a small dataset cannot account for the full natural variability of an, apparently chaotic, process. Although \citeA{Carini2015} provided a much larger field dataset for $Q_b$, these authors did not investigate the cross-shore variation of $Q_b$. Thus, the results presented in Section \ref{qb_results} are novel in the literature and show that $Q_b$ is highly variable at all locations. Such observations are in agreement with previous studies of cross-shore structure of other surf zone parameters, such as $\gamma$ and $H$ \cite{Power2010, Power2015, Martins2017}, and the instantaneous wave speed ($c$) \cite{Postacchini2014, Tissier2015}. Nonetheless, there it was observed a clear connection between the cross-shore structure of $Q_b$ and \citeA{Wright1984} Australian beach morphodynamic model (see Section \ref{res_omega}).

In Section \ref{tides}, a tidal influence on $Q_b$ was observed for the steep beach profile case (Werri Beach). One potential mechanism controlling the changes on $Q_b$'s behaviour was a alteration in the dependence of wave breaking on the local $\gamma$ \cite{Dally1985}. Due to the fact that the profile steepened very quickly in the high tide case (at $x\approx15m$ in Figure \ref{profiles}-f), the waves may not have enough time to adapt to the changes in depth which lead to an increase of shore-breakers. The increase in the occurrence of this breaker type was also confirmed when analysing the raw video imagery. This mechanism seems to be consistent with the fairly constant slope of the $Q_b$ curves regardless of the tidal level. Thus, the rate of change in the amount of wave breaking can be consistent with a lateral expansion of the surf zone with the tidal cycle, and showed a connection between $Q_b$ and beach morphology, which was further explored in Section \ref{res_omega}.

Some other surf zone phenomena were responsible for disturbing the observed patterns in $Q_b$. For instance, infragravity waves were shown to systematically reduce terminal values of $Q_b$ in at least one beach (see Session \ref{ig_energey}). Recent research has shown that infragravity waves modify the water depth in which short waves are propagating, consequently changing the shoreward evolution of sea-swell waves \cite{Tissier2015, Bakker2016a, Padilla2017}. Therefore, if a short-wave sits on the positive part of an infragravity wave propagating shoreward, short-wave breaking may be inhibited because the local water depth ($h$) increases. Such an increase in $h$ causes the local $\gamma$ to decrease, leading the wave to reform or not break \cite{Dally1985} hence lowering $Q_b$.

The infragravity wave control on $Q_b$ could be consistent with break-point infragravity forcing \cite{Symonds1982} or edge waves \cite{Bowen1984, Holman1979, Bowen1978} because these waves can increase water levels asymmetrically, particularly closer to the shore. However, this dynamic is inconsistent with bound wave forcing \cite{LonguetHiggins1964, Battjes2004, Baldock2006} because the net result of the water level changes should be null over a given time interval. Meaning that, in average, the negative part of the incoming bound waves balances the positive part. Therefore, $Q_b$ should remain unchanged in the presence of wave groups. Combining \citeA{Stringari2019} wave tracking and \citeA{DeMoura2017} infragravity wave forcing detection methods could lead to the discovery of the missing link between the two processes and will be attempted in a next publication.

\subsection{Wave Height PDFs}

We have developed novel formulations for broken wave height PDFs that are consistent with observed wave height PDFs (Section \ref{Hbrk_results}). Unfortunately, these results are impossible to directly compare with the existing literature because all previous publications described this distribution as a function of the PDF for all waves \cite{Baldock1998, Battjes1978, Ruessink2003, Alsina2007, Battjes1985, Janssen2007}. Despite this, the overall shape of the distributions shown in Figure \ref{pH_broken_waves} agreed well with TG83's measured broken wave PDFs (the two lower panels in their Figure 10). On a more fundamental level, what these results showed was a direct manifestation of the Central Limit Theorem (CLT) \cite{Feller1945}. Given the systematic sub-sampling of the distribution of all waves ($p(H)$) to obtain  the distribution of broken waves ($p(H_{br})$), it was statistically expected that $p(H_{br})$ be Gaussian shaped, regardless of the shape of its original distribution.

The Weibull PDF was shown to describe $p(H_{br})$ remarkably well, however, the choice for this distribution was based on previous literature rather than on physical or statistical reasoning \cite{Mase1989,Hameed1985,Battjes2000,Mendez2004,Power2016}. Other PDFs, particularly the Gaussian and Gamma distributions, were more accurate than the Weibull PDF to model  $p(H_{br})$ when the K-S test was used to measure their goodness-of-fit. It was found that 97.4\% of the broken wave PDFs were statistically similar to a Gaussian PDF, 96.9\% to a Gamma PDF, 95.7\% to a Weibull PDF, and only 1.4\% were statistically similar to a Rayleigh PDF. The Weibull PDF was then maintained because: 1) the differences seen in the K-S tests were small, and 2) the physical interpretation of the parameters of the alternative distributions are harder to correlate to physical processes (e.g., Gamma PDF), or could allow for negative wave heights (e.g., Gaussian PDF). Unfortunately, $H_{br}$ and $p(H_{br})$ are not usually known, regardless of the distribution used to model them; therefore, Equation \ref{weibull_data} is not of practical use. On the other hand, Equation \ref{true_model} has the potential to lead to a practical model that could provide a more realistic representation of $Q_b$ in energy dissipation models. The approach shown in Section \ref{Hbrk_results} (particularly Equations \ref{transference} and \ref{true_model}) has also the advantage of working with any alternative PDF (e.g., the better performing Gamma and Gaussian PDFs) to describe the evolution of $p(H)$ into $p(H_{br})$.

\subsection{Model performance}\label{model_disc}

The theoretical $Q_b$ models performed poorly when compared to the data from the MLP (see Figure \ref{qb_residual}). Significant improvement was observed for the TG83 model when the Rayleigh PDF was replaced by the Weibull PDF and $n$ was set to 1. Such improvement also followed directly from the CLT whereby this particular combination of parameters made the shape of empirical curve that defines $Q_b$ very close to the curves seen in Figure \ref{pH_broken_waves}. The poor performance of the BJ78 model was due to the truncation of $p(H)$, which directly contradicts the observations presented here (compare Figure \ref{pH_broken_waves} and Figure \ref{qb_plot}) \cite{Baldock1998, Thornton1983, Alsina2007, Janssen2007}.  The reasons why the B98 model performed poorly are unclear, especially when there is strong evidence that its formulation is the best performing in un-calibrated situations \cite{Alsina2007, Apotsos2008, Ruessink2003}. One reason could be that the breaking criterion $H_b$ is not properly describing the start of wave breaking, as previously observed \cite{Ruessink2003}, however, updating $H_b$ to use \citeA{Ruessink2003} formulation did not improve the $Q_b$ residuals for the B98 model (not shown).

Moreover, considering the dissipation term from \citeA{Alsina2007, Janssen2007}, linear wave theory, and conservation of momentum and mass (i.e., there is no reflection at the shoreline, or a swash zone) Equation \ref{E} can be re-written as:

\begin{equation}
  \frac{1}{16} \rho g H^2_{rms} \sqrt{g h} = \frac{3\sqrt{16}} {\pi} B_f \rho g \frac{H^3_{rms}} {h} f Q_b.
\end{equation}

\noindent After algebraic simplifications, it follows that the combination of the remaining variables are of similar order of magnitude. Therefore, even if $Q_b$ is significantly under or over estimated (or even assumed constant), the influence of $Q_b$ on the total energy dissipation can be counter-balanced by optimising either $B_f$, $H_b$, or $\gamma$, as done by  \citeA{Battjes1985, Ruessink2003, Apotsos2008} which is not necessarily a physical improvement to the model.

There exists alternative approaches to model the surf zone energy dissipation which are more realistic than parametric models, e.g., \citeA{Duncan1981} and \citeA{svendsen2006} wave roller models,  \citeA{Smit2013} approach used in SWASH \cite{Zijlema2011} and the recent Smoothed Particle Hydrodynamics (SPH) approach \cite{Altomare2015a}. However, parametric energy dissipation models are still frequently used to drive hydrodinamic \cite{Zhang2018a}, morphological \cite{Larson1990, Hanson1989, Roelvink2009a, DeVriend1993}, and spectral wave models \cite{Booij1999, Ris1999, WW32016}. Therefore, addressing knowledge gaps, such as the ones described in this work,  could result in improvements in these other models which are often used for non-academic coastal management applications.

\subsection{The Problem with Machine Learning}\label{ml_problem}

Machine learning algorithms, particularly neural networks, have revolutionised data problems in the last three decades but have also received intense criticism (e.g., as early as \citeA{Vemuri1993}). Besides the technical implementation and reproducibility problems, there exists a more fundamental issue with the machine learning approach: it does not give new insights into the physical phenomena governing the analysed problem. In this paper, we analysed the feature importance of the input layer and concluded that the MLP was learning from a combination of parameters related to the wave shape, however, the neural connections in the hidden layers are significantly harder to understand. Firstly, because of the enormous number of connections (of the order of  $10^8$); and secondly, because of the non-linearity between these connections. Such intrinsic complexity is a likely cause of the unexpected results shown in Figure \ref{qb_bootstrap}-c), i.e., when the MLP classified data for Moreton Island more accurately when leaving this location out of the training dataset. This result was significantly counter-intuitive given that the practical rules of machine learning state that more training samples result in better accuracy scores \cite{hastie2008}. Several alternatives to the neural network approach were attempted (e.g., logistic regression, Bayesian inference, and nearest neighbours models) but all these methods resulted in accuracy scores of the order of 60\%, which is only slightly better than randomly guessing the wave label.

\section{Conclusion}\label{conclusion}

In this paper, data from seven different Australian beaches across nineteen unique deployments were collected and used to investigate the natural variability in the fraction of broken waves (Qb). A machine learning model that classifies waves into broken or unbroken using wave-by-wave parameters was developed from collocated remote sensing and in-situ pressure transducer data. Using over 350,000 waves classified waves, it was found that Qb is highly variable parameter with a high degree of inter- and intra-beach variability. Nonetheless, correlations between environmental forcing and Qb were found. On steeper beaches, for a given local water depth, Qb was up to 70\% higher at low tide when compared to high tide. In addition, increased infragravity energy levels decreased terminal values of Qb by ~20\%. This correspondence between Qb and environmental parameters was linked to \citeA{Wright1984} beach morphodynamic model: for a given normalised water depth, Qb is higher for dissipative beaches than for intermediate beaches. Using the machine-learning Qb data ($r2>0.99$, $p<<0$) three widely used Qb models were tested and, in general, were shown to perform poorly (average errors of the order of 40\%). The \citeA{Thornton1983} model was significantly improved by replacing the original Rayleigh PDF with a Weibull PDF (average errors $<$ 10\%). Finally, a mathematically sound transformation of the PDF for all wave heights $(p(H))$ into the PDF of broken wave heights $(p(H_{br}))$ was outlined based on the patterns of p(Hbr) observed here. The novel Qb dataset derived here, shows that the current theoretical parameterizations for Qb are poor predictors because they cannot account for the full natural variability of the parameter. This dataset is used to develop a novel, data-driven, method to transform $p(H)$ into $p(H_{br})$ that could be used to further improve coastal management tools.

%%%%%%%%%%%%%%%%%%%%%%%%%%%%%%%%%%%%%%%%%%%%%%%%%%%%%%%%%%%%%%%%%
%%
%%  ACKNOWLEDGMENTS
%%
%% The acknowledgments must list:
%%
%% >>>>	A statement that indicates to the reader where the data
%% 	supporting the conclusions can be obtained (for example, in the
%% 	references, tables, supporting information, and other databases).
%%
%% 	All funding sources related to this work from all authors
%%
%% 	Any real or perceived financial conflicts of interests for any
%%	author
%%
%% 	Other affiliations for any author that may be perceived as
%% 	having a conflict of interest with respect to the results of this
%% 	paper.
%%
%%
%% It is also the appropriate place to thank colleagues and other contributors.
%% AGU does not normally allow dedications.
%
%
\acknowledgments

The field work (2014 experiments) were funded by a University of Newcastle Faculty of Science and I.T. Strategic Initiatives Research Fund Grant 2014 to HEP. The authors are grateful to Alex Atkinson, Andrew Magee, Annette Burke, Emily Kirk, Daniel Harris, David Hanslow, Kaya Wilson, Madeleine Broadfoot, Michael Hughes, Mike Kinsela, Murray Kendall, Rachael Grant, Rebecca Hamilton, Samantha Clarke, and Tom Donaldson-Brown who assisted with the field data collection. Caio E. Stringari is funded by a University of Newcastle Research Degree Scholarship (UNRS) 5050UNRS and a Central \& Faculty scholarship. The pressure transducer and video cameras used in the 2014 experiments were kindly lent by Tom Baldock from the University of Queensland. The sediment data used in Section \ref{res_omega} were kindly provided by professor Andrew Short from the University of Sydney. The authors are also thankful to the Academic Research Computing Support Team, particularly Aaron Scott, at the University of Newcastle for support with the I.T. infrastructure on which all video data pre-processing and machine-learning development were undertaken, and to Bas Hoonhout who helped providing the original image rectification codes.

\section*{Data availability}

The code, data, and the pre-trained neural network used in this work will available at \url{https://github.com/caiostringari/pywavelearn} after this manuscript is published.

%
%% ------------------------------------------------------------------------ %%
%% References and Citations

%%%%%%%%%%%%%%%%%%%%%%%%%%%%%%%%%%%%%%%%%%%%%%%
% BibTeX is preferred:
%
% \bibliography{/home/stringari/DocHub/Bibliography/library.bib}
%
% don't specify bibliographystyle
%%%%%%%%%%%%%%%%%%%%%%%%%%%%%%%%%%%%%%%%%%%%%%%

% Please use ONLY \citeA and \cite for reference citations.
% DO NOT use other cite commands (e.g., \cite, \citeyear, \nocite, \citealp, etc.).
%% Example \citeA and \cite:
%  ...as shown by \citeA{Boug10}, \citeA{Buiz07}, \citeA{Fra10},
%  \citeA{Ghel00}, and \citeA{Leit74}.

%  ...as shown by \cite{Boug10}, \cite{Buiz07}, \cite{Fra10},
%  \cite{Ghel00, Leit74}.

%  ...has been shown \cite [e.g.,][]{Boug10,Buiz07,Fra10}.

\newpage
\section*{Appendix 1}\label{timexes}

\begin{figure}[htp]
	\centering
	\includegraphics[scale=0.6]{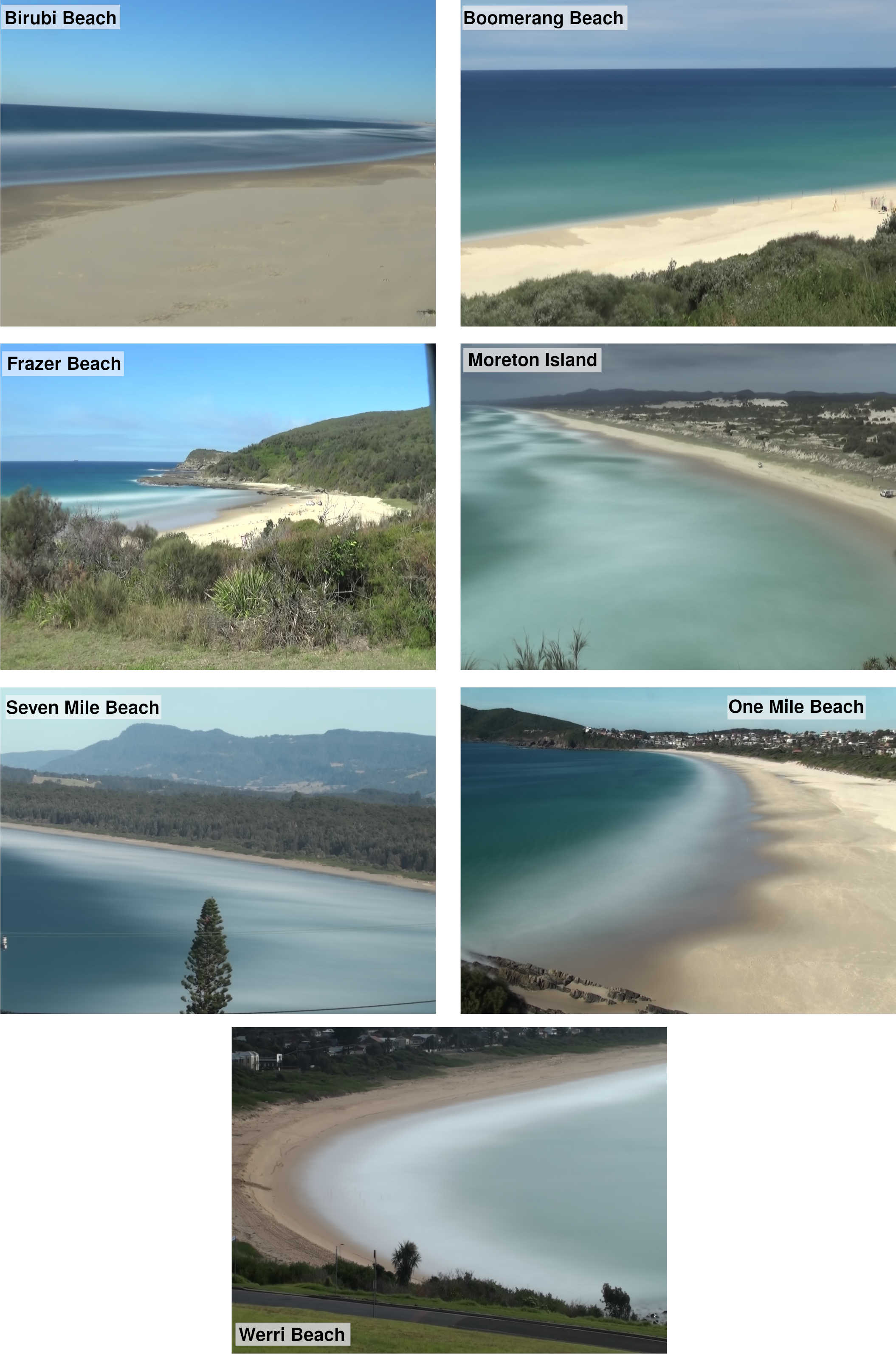}
	\caption{Time average images (Timex) using 10 minutes of data for all locations analysed.}\label{timex}
\end{figure}

\end{document}